\documentclass[journal]{IEEEtran}
\IEEEoverridecommandlockouts
\usepackage{graphicx,amssymb,amsmath,color}
\usepackage{epstopdf}
\usepackage[mathscr]{euscript}
\usepackage{dsfont}
\usepackage[noadjust]{cite}
\usepackage[font=footnotesize]{caption}
\usepackage[font=footnotesize]{subcaption}
\usepackage[labelsep=period]{caption}
\usepackage{bm} 
\usepackage{multirow} 
\usepackage{psfrag}
\usepackage{array,booktabs}
\usepackage{txfonts}
\usepackage{lipsum}
\usepackage{float}
\usepackage{enumerate}
\usepackage{enumitem}
\usepackage[english]{babel}
\usepackage{blindtext}
\usepackage{mathtools}
\usepackage{relsize}
\usepackage[subnum]{cases}
\usepackage[flushleft]{threeparttable}
\usepackage[hyphens]{url}
\usepackage{algorithmic}
\usepackage{algorithm}
\usepackage[none]{hyphenat}
\usepackage{hyperref}

\newcommand{\cPsi}{{\cos \Psi}}

\newcommand{\EX}[1]{{\mathbb{E}}\left[{#1}\right]}
\newcommand{\EXs}[2]{{\mathbb{E}}_{{#1}~}\!\!\left[{#2}\right]}

\newcommand{\PX}[1]{{\mathbb{P}}\left[{#1}\right]}

\newcommand{\ssin}[1]{\sin\left( {#1} \right)}

\newcommand{\locparam}[1]{\mu_{\text{{#1}}}}
\newcommand{\scaleparam}[1]{b_{\text{{#1}}}}

\DeclareMathOperator*{\argmax}{arg\,max}

\makeatletter
\newcommand{\vast}{\bBigg@{4}}
\newcommand{\Vast}{\bBigg@{5}}
\makeatother

\newcounter{tempEquationCounter} 
\newcounter{thisEquationNumber}
\newenvironment{floatEq}
{\setcounter{thisEquationNumber}{\value{equation}}\addtocounter{equation}{1}
	\begin{figure*}[!t]
		\normalsize\setcounter{tempEquationCounter}{\value{equation}}
		\setcounter{equation}{\value{thisEquationNumber}}
	}
	{\setcounter{equation}{\value{tempEquationCounter}}
		\hrulefill\vspace*{4pt}
	\end{figure*}
}

\newcommand*{\QEDB}{\hfill\ensuremath{\blacksquare}}%
\newenvironment{proof}[1][Proof:]{\begin{trivlist}
		\item[\hskip \labelsep {\itshape\bfseries #1}]}{\QEDB \end{trivlist}}

\newenvironment{proposition}[1][Proposition]{\begin{trivlist}
		\item[\hskip \labelsep {\itshape\bfseries #1}]}{\end{trivlist}}

\addto\captionsenglish{\renewcommand{\figurename}{Fig.}}

\begin{document}
	% \title{On the Random Orientation of LiFi-Enabled Devices: Experiments, Analysis, and Applications}
	\vspace{-1cm}
	\title{Modeling the Random Orientation of Mobile Devices: Measurement, Analysis and LiFi Use Case}
	
	\author{
		\IEEEauthorblockN{Mohammad~Dehghani~Soltani,~\IEEEmembership{Student Member,~IEEE,} Ardimas~Andi~Purwita,~\IEEEmembership{Student Member,~IEEE,} Zhihong~Zeng,~\IEEEmembership{Student Member,~IEEE,} Harald~Haas,~\IEEEmembership{Fellow,~IEEE} and Majid~Safari,~\IEEEmembership{Member,~IEEE} }\\
		%\IEEEauthorblockA{LiFi R\&D Center, Institute for Digital Communications, The University of Edinburgh, Edinburgh, EH9 3JL, UK\\
		%Email: \{m.dehghani, a.purwita, zhihong.zeng, h.haas, majid.safari\}@ed.ac.uk}
		\thanks{
			The authors are with the LiFi Research and Development Centre, Institute for Digital Communications, The University of Edinburgh, EH9 3JL, UK. (e-mail: \{m.dehghani, a.purwita, zhihong.zeng, h.haas, majid.safari\}@ed.ac.uk).}
		\vspace{-0.85cm}
	}

	\maketitle

	\begin{abstract}
		Light-fidelity (LiFi) is a networked optical wireless communication (OWC) solution for high-speed indoor connectivity for fixed and mobile optical communications. Unlike conventional radio frequency wireless systems,  the OWC channel is not isotropic, meaning that the device orientation affects the channel gain significantly, particularly for mobile users. However, due to the lack of a proper model for device orientation, many studies have assumed that the receiver is vertically upward and fixed. In this paper, a novel model for device orientation based on experimental measurements of forty participants has been proposed. 
% * <majid.safari@ed.ac.uk> 2018-05-14T10:28:36.882Z:
% 
% > forty
% is it exactly forty or can we say more than forty?
% 
% ^.
		It is shown that the probability density function (PDF) of the polar angle can be modeled either based on a Laplace (for static users) or a Gaussian (for mobile users) distribution. In addition, a closed-form expression is obtained for the PDF of the cosine of the incidence angle based on which line-of-sight (LOS) channel gain is described in OWC channels. An approximation of this PDF based on the truncated Laplace is proposed and the accuracy of this approximation is confirmed by the Kolmogorov-Smirnov distance (KSD). 
		Moreover, the statistics of the LOS channel gain are calculated and the random orientation of a user equipment (UE) is modeled as a random process. The influence of the random orientation on signal-to-noise-ratio (SNR) performance of OWC systems has been evaluated. Finally, an orientation-based random waypoint (ORWP) mobility model is proposed by considering the random orientation of the UE during the user's movement. The performance of ORWP is assessed on the handover rate and it is shown that it is important to take the random orientation into account. 
	\end{abstract}
	
	\vspace{-0.3cm}
	
	\section{Introduction}
	\IEEEPARstart{T}{he} increasing request for wireless data, which is expected to be $49$ exabytes per month by $2021$ \cite{Cisco}, motivates both academia and industry to invest in alternative solutions. These include mmWave, massive multiple-input multiple-output (MIMO), free space optical communication and Light-Fidelity (LiFi) to support the data traffic growth  and next-generation high-speed wireless communication systems. 
	Among these technologies, LiFi is a novel bidirectional, high-speed and fully networked wireless communication technology. LiFi uses visible light as the propagation medium in the downlink for the purposes of illumination and communication. It can use infrared in the uplink so that the illumination constraint of a room remains unaffected, and also to avoid interference with the visible light in the downlink \cite{Haas,MDSFeedback}. %It has been experimentally shown that $3.46$ Gb/s data rate over $5$ m free-space link can be achieved using visible light communication (VLC) with micro light emitting diode (LED) \cite{Islim} which is very promising for future indoor communications. 
	LiFi offers a number of important benefits that have made it favorable for recent and future research. These include the very large, unregulated bandwidth available in the visible light spectrum (more than $10^3$ times greater than the whole RF spectrum), high energy efficiency \cite{ImanICT2018}, the rather straightforward deployment which uses off-the-shelf light emitting diodes (LED) and photodiode (PD) devices at the transmitter and receiver ends respectively, and enhanced security as light does not penetrate through opaque objects \cite{WuVLC5G}. One of the key shortcomings of the current research literature on LiFi is the lack of appropriate statistics of device orientation and rotation modeling for system design and handover management purposes.

Smartphones are the most significant and indispensable part of the wireless network generating about $86\%$ of the mobile data traffic \cite{Cisco}. LiFi as part of future $5$G can handle this immense data traffic thanks to future LiFi-enabled smartphones. Generally, users tend to work with their smartphones in a comfortable manner which is not necessarily vertically upward. Smartphones are equipped with a gyroscope that can measure the device orientation. This orientation information can be fedback to the access point (AP) via limited-feedback methods \cite{MDSFeedback,dehghani2015limited,soltani2017throughput}. Then, the AP can use the orientation information for resource allocation or handover management.
Many previous studies assumed that the receiver is vertically upward and fixed for simplicity purposes and also due to the lack of a proper model for device orientation in LiFi networks. However, there are few studies that have considered the effect of device orientation in their analysis \cite{APselection,MDSHandover,ArdimasVTC2018, ImpactTilted,ICCTilting,MDSBER,huynh2016vlc, LixuanPositioning2017,EricPositioning,jeong2013tilted,wang2011performance, ErogluArxiv2017Orientation}. Nevertheless, none of these studies have considered the actual statistics of device orientation and have mainly assumed uniform or Gaussian distribution with hypothetical moments for device orientation. 
	%All these studies consider the polar angle as a random variable without taking into account the actual statistics of it, such as its mean and variance. The paper at hand addresses this issue.
	In this study, based on practical measurements from forty participants, a new statistical model for device orientation is proposed. %The statistics of this model has been derived and presented. 
	
	%Smart cell phones are equipped with an accelerometer, gyroscope and compass that enable them to measure the instantaneous orientation of the device in three dimensions. This can be done by reporting the rotation about each axes. These rotation angles are known as the Euler rotation angles \cite{palais2009disorienting}.  

	%Optical wireless communications (OWCs) can be considered as complementary to radio frequency (RF) for wireless access, and is attractive for high-speed short-range links in settings where RF is not desired or restricted. 

	\vspace{-5pt}
	\subsection{Literature Review and Motivation}
	
	In \cite{APselection}, the authors consider three standard angles similar to those used in mobile devices to model the device orientation; namely yaw, pitch and roll. Based on this model, the effect of arbitrary orientation on users' throughput and network load balancing is investigated. 
	The problem of handover due to device rotation for downlink in an indoor
	optical attocell network was first proposed in \cite{MDSHandover}. The handover probability has been obtained for both sitting and mobile users while considering device orientation. In \cite{ArdimasVTC2018}, the handover probability in hybrid LiFi/RF-based networks is evaluated assuming randomly-oriented devices. 
	The impact of the receiver's tilted angle on the channel capacity of visible
	light communications (VLCs) is investigated in \cite{ImpactTilted}. The lower and upper bounds of the channel capacity for the VLC are presented and by considering an optimization problem the channel capacity has been improved by tilting the receiver plane properly. 
	In \cite{ICCTilting}, a theoretical expression of the bit error ratio (BER) for input-dependent noise of VLC using on-off keying has been derived. Then, a convex optimization problem is formulated based on the derived BER expression to minimize the BER performance by tilting the receiver plane.  The impact of device orientation on BER performance of DC biased optical orthogonal frequency division multiplexing (DCO-OFDM) has been evaluated in \cite{MDSBER}. A closed form approximation for BER of randomly-orientated UEs is derived.
%	The impact of device orientation has been evaluated on BER performance of DC biased optical orthogonal frequency division multiplexing (DCO-OFDM) in \cite{MDSBER}. By introducing two elevation angles called critical and best aligned angles, the significance roles of them are shown on BER performance in LiFi networks.  
	
	The effect of device orientation on positioning has been investigated in several studies \cite{huynh2016vlc, LixuanPositioning2017, EricPositioning, jeong2013tilted}. By using the accelerometer sensor, the positioning technique proposed in \cite{huynh2016vlc} can be used for any arbitrary device orientation. Downlink and uplink indoor positioning techniques based on VLC while considering the tilting of the device have been developed in \cite{LixuanPositioning2017} and \cite{EricPositioning}, respectively. It is shown that the tilting angle can affect the positioning error significantly. Therefore, device orientation should be considered in the positioning analysis \cite{jeong2013tilted}. The signal-to-noise-ratio (SNR) and spectral efficiency improvement of OFDM signals for indoor visible light communication by optimally tilting the receiver plane is proposed in \cite{wang2011performance}. 
	In \cite{ErogluArxiv2017Orientation}, the effect of random orientation on the line-of-sight (LOS) channel gain for a randomly located user equipment (UE) has been investigated. The statistical distribution of the channel gain has been derived for a single LED and extended to a scenario with double LEDs.
	We note that  none of  these  studies  are  supported  by  any  experimental  data.
	A measurement of the random orientations of mobile devices has been  made  in  \cite{peng2017three},  but  the  authors  only  measure  the statistics of the pace of change of the device orientation.  Their results, therefore, do not describe the  statistical model of the randomly-oriented devices in general. 
	Our recent work in \cite{ArdimasWCNC2018} reports some initial results based on the experimental data from $40$ participants.
	%Recent work has been reported in \cite{ArdimasWCNC2018}, which is based on the experimental results taken from $40$ participants. 
	The effect of randomly-oriented devices has been studied and a statistical model of the LiFi channel considering the random orientation is proposed. %The authors show that the distribution of the cosine of the incidence angle follows a Laplacian form. 

	%\vspace{-0.3cm}
	\subsection{Contributions and Outcomes}
	The lack of a proper model for device orientation along with its analysis was a motivation to perform a set of experimental measurements, in which participants use their smartphones in landscape or portrait mode in both sitting and walking conditions. It is worth mentioning that the initial statistical model proposed for device orientation can be also used in mmWave communications as it depends on the user behavior and not the access technology. The main contributions of this paper are summarized as follows:
	
	$\bullet$ Performing a set of practical measurements, proposing a new model for device orientation based on the measurement results and deriving the probability density function (PDF) and statistics of the model. 
	
	$\bullet$ Deriving the PDF and statistics of both LOS channel gain and received SNR.
	
	%$\bullet$ Modeling the effect of the random orientation of UE as slow fading and evaluating its impact on BER performance of optical wireless communications.  
	
	$\bullet$ Proposing an orientation-based random waypoint (ORWP) mobility model that considers the random orientation of the UE while the user is moving. The ORWP mobility model can also be used in mmWave networks due to the importance of the angle of arrival in those systems.

	%\subsection{Paper Organization}

	\textit{Notations:} $f_X$ and $F_X$ denote the PDF and the cumulative distribution function (CDF) of the random variable (RV) $X$, respectively. $\EX{\cdot}$ denotes the expected value over the RV inside the argument, and $\EXs{X}{\cdot}$ denotes the expected value with respect to the RV, $X$. $[]^{\rm{T}}$ stands for transpose operator. The inner product operator is shown by $\cdot$ and $\lVert\cdot\rVert$ expresses the norm of a vector. Further, $\tan^{-1}(y/x)$ is the four-quadrant inverse tangent. 
	% %%%%%%%%%%%%%%%%%%%%%%%%%%%%%%%%%%%%%%%%%%%%%%%%%%%%%%%%%%%%%%%%%%%%%%%
	% %%%%%%%%%%%%%%%%%%%%%%%%%%%%%%%%%%%%%%%%%%%%%%%%%%%%%%%%%%%%%%%%%%%%%%%
	% %%%%%%%%%%%%%%%%%%%%%%%%%%%%%%%%%%%%%%%%%%%%%%%%%%%%%%%%%%%%%%%%%%%%%%%
	\section{Rotation Geometry}
	
	%In this paper, we are interested in the orientation of UE in terms of $\ccos{\psi}$ since the path loss model for the optical wireless communications is proportional to $\ccos{\psi}$ under the assumption of the Lambertian propagation model.
	
\begin{figure}
		\centering
		\begin{subfigure}[b]{0.5\columnwidth}
			\centering
			\includegraphics[width=1\columnwidth,draft=false]{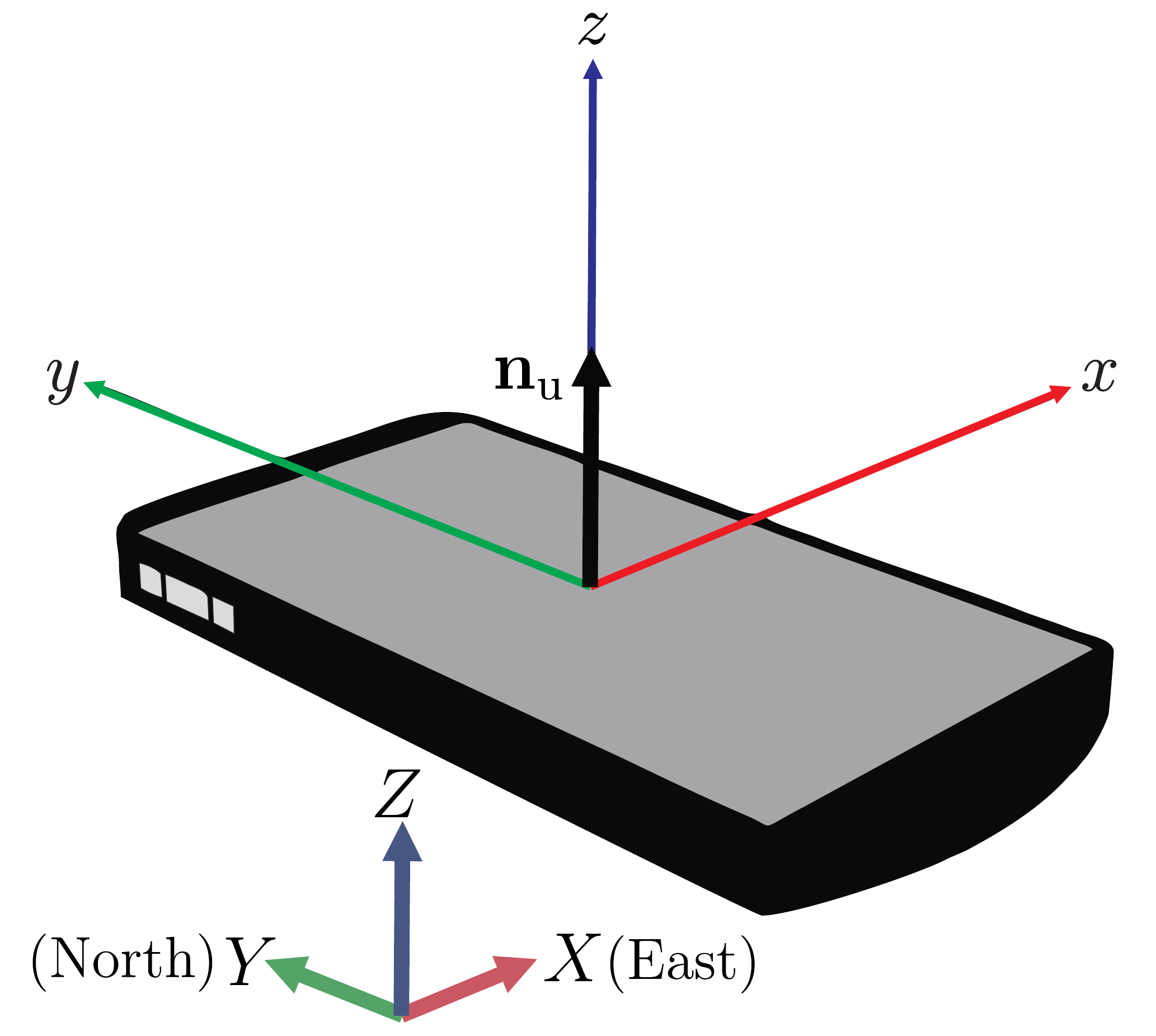}
			\caption{}
		\end{subfigure}%
		~
		\begin{subfigure}[b]{0.5\columnwidth}
			\centering
			\includegraphics[width=1\columnwidth,draft=false]{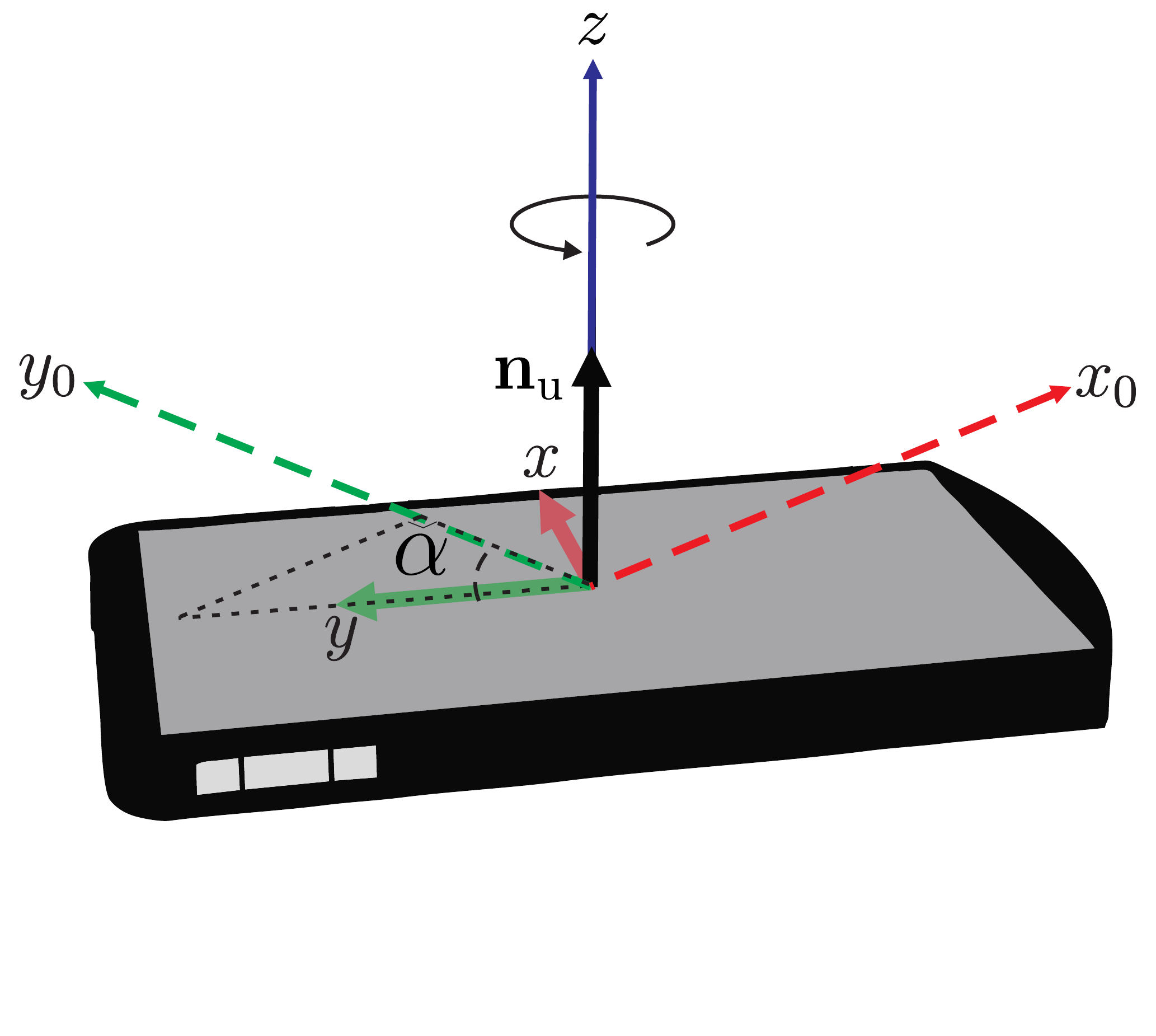}
			\caption{}
		\end{subfigure}\\
		\begin{subfigure}[b]{0.5\columnwidth}
			\centering
			\includegraphics[width=1\columnwidth,draft=false]{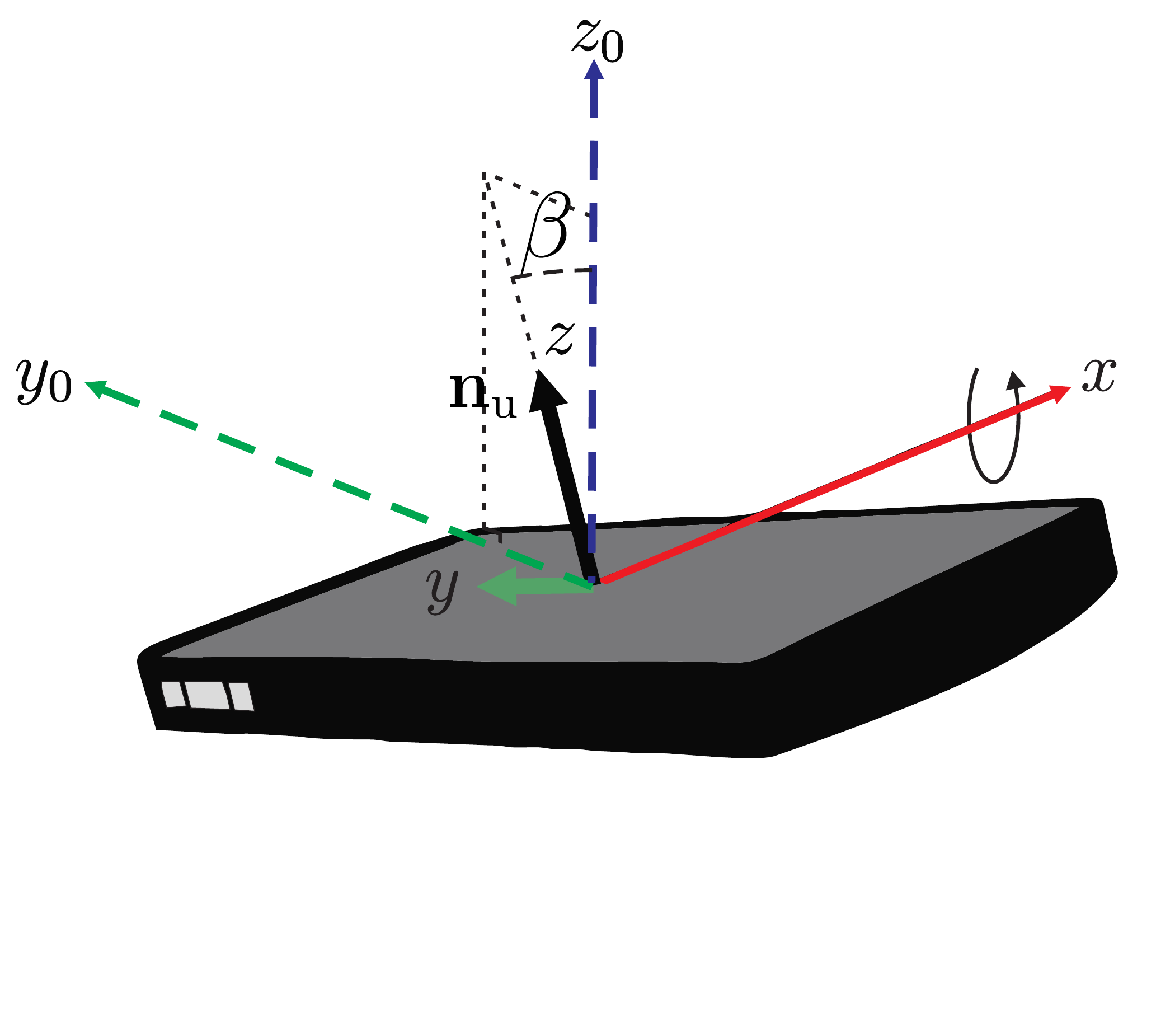}\vspace{-0.2cm}
			\caption{}
		\end{subfigure}%
		~
		\begin{subfigure}[b]{0.5\columnwidth}
			\centering
			\includegraphics[width=1\columnwidth,draft=false]{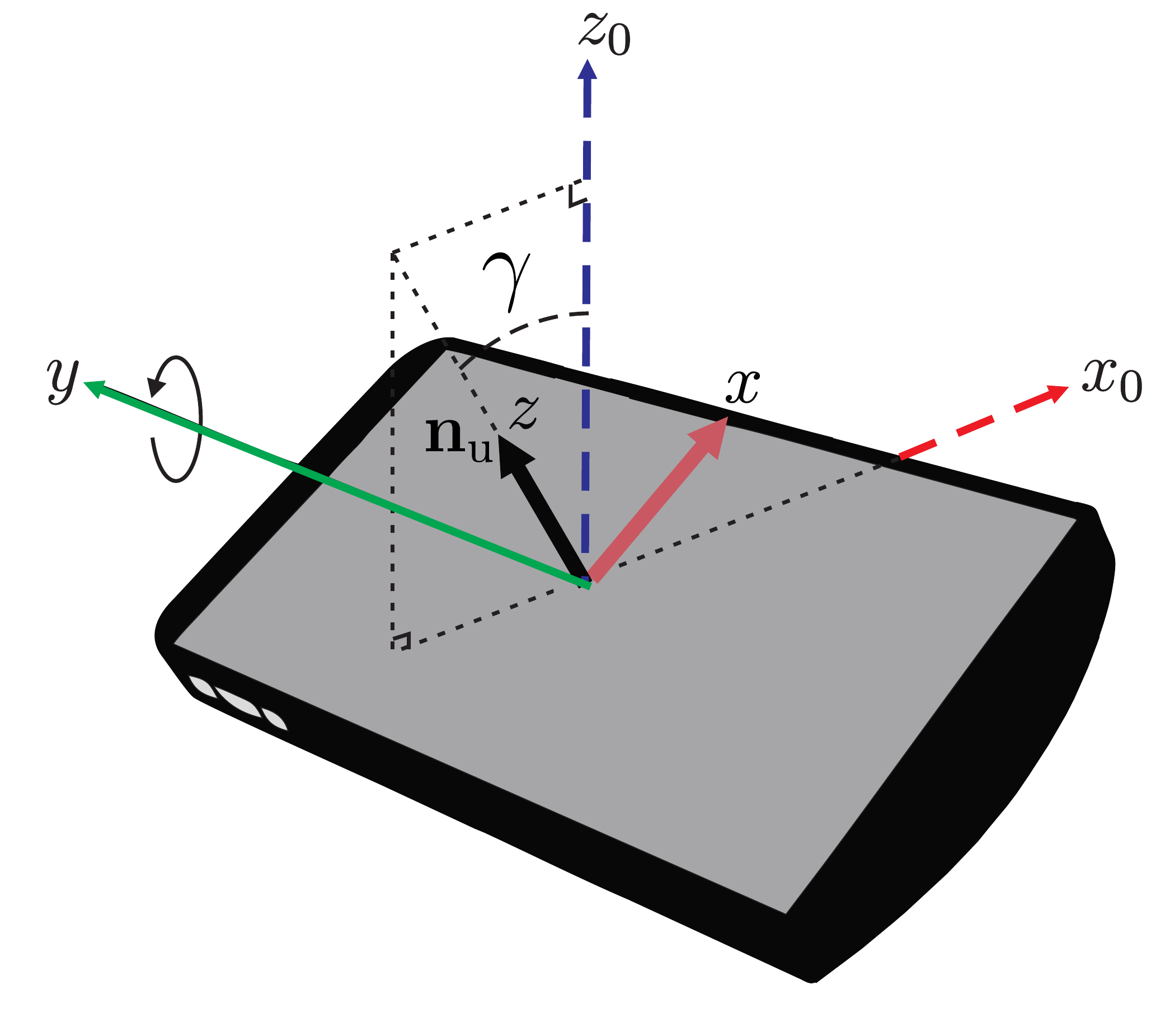}\vspace{-0.2cm}
			\caption{}
		\end{subfigure}
		\caption{Orientations of a mobile device: (a) normal position, (b) yaw rotation with angle $\alpha$, about the $z$-axis (c) pitch rotation with angle $\beta$, about the $x$-axis and (d) roll rotation with angle $\gamma$, about the $y$-axis.}
		\label{figori}
		\vspace{-10pt}
	\end{figure}

	%\subsection{Rotations Representation}
	According to Euler's rotation theorem \cite{kuipers1999quaternions}, any rotation in $\mathbb{R}^3$ space can be uniquely achieved by composing three elemental rotations, i.e., the rotations about the axes of a coordinate system. Depending on whether the device (local) or Earth (global) coordinate system is chosen, there are two types of rotations. Intrinsic rotation corresponds to a rotation about the device coordinate system and extrinsic rotation, which conforms to a rotation about the Earth coordinate system. Throughout this paper, we will show the device and Earth coordinate system with $xyz$ and $XYZ$, respectively. The Earth and device coordinates are shown in Fig.~\ref{figori}(a).

	%According to the Euler's rotation theorem, any rotation in the three-dimensional space of $\mathbb{R}^3$ can be uniquely identified by two characteristics: i) the rotation axis, which is a one-dimensional subspace of $\mathbb{R}^3$ ; and ii) the plane orthogonal to the axis of rotation, which is a two-dimensional subspace of $\mathbb{R}^3$ \cite{MDSHandover}. Therefore, any orientation in $\mathbb{R}^3$ can be achieved by composing three elemental rotations, which are rotations about the axes of a coordinate system. Euler angles are defined by three of these rotations. 
	
	Since the matrix multiplication is not commutative, the elemental rotation order matters. Therefore, there are six possible choices of rotation axes for proper Euler angles. 
	World wide web consortium (W3C) has chosen the intrinsic rotation orders of $(z\rightarrow x' \rightarrow y'')$ as standard for device orientation where $x'y'z'$ and $x''y''z''$ are respectively the device coordinate systems after rotation about the $z$-axis and subsequently after rotation about  $x'$-axis \cite{W3C2016}. These rotation orders will be also adopted in this study. According to W3C specification, the angles $\alpha \in [0, 360)$, $\beta\in[-180,180)$ and $\gamma\in[-90,90)$ correspond to the rotation about $z$, $x$ and $y$ axes, respectively. The elemental rotation angles, $\alpha, \beta$ and $\gamma$ are called yaw, pitch and roll, respectively. The rotation about the axes are depicted in Fig.~\ref{figori}(b)-(d).
	
	\begin{floatEq}
		\begin{align} \label{eqrotationmatrix}
		\mathbf{n}'_{\rm{u}} = 
		\mathbf{R}_\alpha \mathbf{R}_\beta \mathbf{R}_\gamma
		\begin{bmatrix}
		0 \\
		0 \\
		1 
		\end{bmatrix}
		&=
		\begin{bmatrix}
		\cos \alpha & -\sin \alpha & 0 \\
		\sin \alpha & \cos \alpha & 0 \\
		0 & 0 & 1 
		\end{bmatrix}
		\begin{bmatrix}
		1 & 0 & 0 \\
		0 & \cos \beta & -\sin \beta \\
		0 & \sin \beta & \cos \beta 
		\end{bmatrix}
		\begin{bmatrix}
		\cos \gamma & 0 & \sin \gamma \\
		0 & 1 & 0 \\
		-\sin \gamma & 0 & \cos \gamma 
		\end{bmatrix} 
		\begin{bmatrix}
		0 \\
		0 \\
		1 
		\end{bmatrix}
		= 
		\begin{bmatrix}
		\cos \gamma \sin \alpha \sin \beta+\cos \alpha \sin \gamma \\
		\sin \alpha \sin \gamma-\cos \alpha \cos \gamma \sin \beta \\
		\cos \beta \cos \gamma 
		\end{bmatrix}.
		\end{align} 
		\vspace{-15pt}
	\end{floatEq}

	Now we derive the concatenated rotation matrix with respect to the Earth coordinate system. Let $\mathbf{n}_{\rm{u}}=[n_1, n_2, n_3]^{\rm{T}}$ and $\mathbf{n}'_{\rm{u}}=[n'_1, n'_2, n'_3]^{\rm{T}}$ be the normal vectors before and after device rotation. Based on the Euler's theorem, we have $\mathbf{n}'_{\rm{u}}=\mathbf{R}_\alpha\mathbf{R}_\beta\mathbf{R}_\gamma \mathbf{n}_{\rm{u}}$, where $\mathbf{R}_\alpha$, $\mathbf{R}_\beta$, and $\mathbf{R}_\gamma$ denote the rotation matrices with respect to the $z$, $x$ and $y$ axes, respectively. Let's $\mathbf{R}=\mathbf{R}_\alpha\mathbf{R}_\beta\mathbf{R}_\gamma$ be the ultimate rotation matrix. Assuming that the Earth and device coordinate systems initially coincide and $\mathbf{n}_{\rm{u}}=[0, 0, 1]^{\rm{T}}$, the corresponding rotated normal vector, after applying the rotation matrices $\mathbf{R}_\alpha$, $\mathbf{R}_\beta$, and $\mathbf{R}_\gamma$, is given by (\ref{eqrotationmatrix}) at the top of this page. Note that since the initial device coordinates are aligned with the Earth coordinates, the rotation matrix, $\mathbf{R}$, can be represented in the Earth coordinates, $XYZ$.
	As can be seen, the rotated normal vector is a function of the three elemental Euler's angles. 
	
	%The normal vector $\mathbf{n}_{\rm{u}}$ can be represented by the angles $\alpha, \beta$, and $\gamma$ as expressed in (\ref{eqrotationmatrix}) shown on top of the next page, where $\mathbf{R}_\alpha$, $\mathbf{R}_\beta$, and $\mathbf{R}_\gamma$ denote the rotation matrices w.r.t. $\alpha$, $\beta$, and $\gamma$, respectively. 

	The rotated normal vector, $\mathbf{n}'_{\rm{u}}$, can be represented in the spherical coordinate system (corresponding to $XYZ$) with the polar angle $\theta$ and azimuth angle $\omega$. Thus, $\theta$ is the angle between $\mathbf{n}'_{\rm{u}}$ and the positive direction of the $Z$-axis, while $\omega$ denotes the angle between the projection of $\mathbf{n}'_{\rm{u}}$ in the $XY$-plane and positive direction of the $X$-axis. Note that $\cos\theta=\mathbf{n}'_{\rm{u}}\cdot\hat{Z}/\lVert\mathbf{n}'_{\rm{u}} \rVert$, where $\hat{Z}=[0,0,1]^{\rm{T}}$ is the unit vector of the $Z$-axis. Then, from \eqref{eqrotationmatrix}, the polar angle $\theta$ can be obtained as:\vspace{-0.1cm}
	\begin{align}\label{eqtheta}
	\theta = \cos^{-1} \left( \cos\beta \cos\gamma \right).
	\end{align}
	It is clear from (\ref{eqtheta}) that the polar angle only depends on the pitch and roll rotation angles which are further associated with the movements of human's wrists.
	
	Referring to Fig.~\ref{figsphericalcoordinate}, the azimuth angle, $\omega$, can be obtained as follows:\vspace{-0.0cm}
	\begin{equation}
	\label{omegaeq}
	\omega=\tan^{-1}\left(\frac{n'_2}{n'_1} \right)= \tan^{-1}\left(\frac{\sin \alpha \sin \gamma-\cos \alpha \cos \gamma \sin \beta}{\cos \gamma \sin \alpha \sin \beta+\cos \alpha \sin \gamma} \right) 
	\end{equation}
	
	From a mobility point of view, let's define the angle $\Omega$ which represents the angle between the direction of movement and the $X$-axis in the Earth coordinate system. The angle $\Omega$ can be described as the angle between  
	the projection of the $y$-axis on the $XY$-plane and $X$-axis when the user is working with the cellphone in portrait mode, or the angle between the projection of the $x$-axis on the $XY$-plane and $X$-axis when the user is working with the cellphone in landscape mode. Thus,\vspace{-0.0cm}
	\begin{equation}
	\Omega=\begin{cases}
	\cos^{-1}\left(\mathbf{R}\hat{y}\cdot\hat{X} \right) \ \ \ \ \ \ \ \ \  {\rm{Portrait\ mode}} \\
	\cos^{-1}\left(\mathbf{R}\hat{x}\cdot\hat{X} \right) \ \ \ \ \ \ \ \ \  {\rm{Landscape\ mode}}
	\end{cases},
	\end{equation}
	where $\hat{y}$ and $\hat{x}$ are the unit vectors of $y$ and $x$ axes in the device coordinate system and $\hat{X}$ is the unit vector of $X$-axis in the Earth coordinate system. When the user is working with the cellphone in portrait mode, the angle between the $y$-axis and North is defined as $\alpha$, and $\Omega$ is specified based on the angle between the $y$-axis and East. Hence, the relationship between $\Omega$ and $\alpha$ can be expressed as:
	\begin{equation}
	\label{OmegaPortrait}
	\Omega=\begin{cases}
	\alpha+\frac{\pi}{2} \ \ \ \ \ \ \ \ \ \ {\alpha\in(0,\frac{3\pi}{2}]} \\\vspace{-0.1cm}
	\alpha-\frac{3\pi}{2} \ \ \ \ \ \ \ \ \  {\alpha\in(\frac{3\pi}{2},2\pi]}
	\end{cases}.
	%\vspace{-5pt}
	\end{equation}
	Since the difference between portrait mode and landscape mode is just $\pi/2$ which follows the right-hand rule, for landscape mode, we have:\vspace{-0.2cm}
	\begin{equation}
	\label{OmegaLandscape}
	\Omega=\begin{cases}
	\alpha+\pi \ \ \ \ \ \ \ \ \ \ \ \ \ \ {\alpha\in(0,\pi]} \\\vspace{-0.1cm}
	\alpha-\pi \ \ \ \ \ \ \ \ \ \ \ \ \ \  {\alpha\in(\pi,2\pi]}
	\end{cases}.
	\vspace{-5pt}
	\end{equation}

	%In $\mathbb{R}^3$ space, it is typical to represent a vector in spherical coordinates, i.e., a polar angle denoted by $\theta \in [0, \pi]$ and an azimuth angle denoted by $\omega \in [0, 2\pi)$ as depicted in Fig.~\ref{figsphericalcoordinate}. 
	
	% \begin{figure}
	%     \begin{center}
	%         \includegraphics[width=0.6\columnwidth,draft=false]{./figs/sperhicalcoordinates.eps}
	%         \caption{Spherical coordinates representation.}
	%         \label{figsphericalcoordinate}
	%     \end{center}
	% \end{figure}
	
	%In this paper, we are interested in statistical model of the polar angle $\theta$ which depends on the movement of human's wrists. Let $\mathbf{n}_{\rm{u}}$ be represented by a column vector $[n_{{\rm{u,x}}}, n_{{\rm{u,y}}}, n_{{\rm{u,z}}}]^{\text{T}}$. From (\ref{eqrotationmatrix}) and $\cos(\theta) = n_{{\rm{u,z}}}/\lVert\mathbf{n}_{\rm{u}} \rVert = n_{{\rm{u,z}}}$, the spherical coordinates can be written as:
	%\begin{align}\label{eqtheta}
	%	\theta = \cos^{-1} \left( \cos\left(\beta\right) \cos\left(\gamma\right) \right).
	%\end{align}
	
	%It is clear from (\ref{eqtheta}) that the polar angle highly depends on the pitch and roll rotations which further depend on the movements of human's wrists. Throughout this paper, the azimuth angle, $\omega$, is known.

	% %%%%%%%%%%%%%%%%%%%%%%%%%%%%%%%%%%%%%%%%%%%%%%%%%%%%%%%%%%%%%%%%%%%%%%%
	% %%%%%%%%%%%%%%%%%%%%%%%%%%%%%%%%%%%%%%%%%%%%%%%%%%%%%%%%%%%%%%%%%%%%%%%
	% %%%%%%%%%%%%%%%%%%%%%%%%%%%%%%%%%%%%%%%%%%%%%%%%%%%%%%%%%%%%%%%%%%%%%%%
	\section{Mobile Device Orientation Statistics }
	\label{ExperimentalSetup}
	An experiment is designed to study mobile users behavior and to  develop a statistical model for the orientation of mobile devices that act as the receiver for wireless communication systems. During the experiment, $40$ participants were asked to use their cellphones normally that create $222$ datasets for orientation. They were asked to use the cellphone in both portrait and landscape modes for one minute. The orientation data is measured for both sitting and mobile users. In the experimental measurement, the application Physics Toolbox Sensor Suite has been used as it can provide instantaneous rotation angles, $\alpha$, $\beta$ and $\gamma$ \cite{androidapp}. 
	This application can be running in the background while the participants can perform activities that require data connection, e.g., browsing or watching streaming videos. Below is a summary of the experimental setup:
	
	%In the experimental measurement, the application Physics Toolbox Sensor Suite has been used so that it can provide instantaneous rotation angles, $\alpha$, $\beta$ and $\gamma$ \cite{androidapp}. 
	%This application can be running in background while the participants can perform activities that require data connection, e.g., browsing or watching streaming video. It is observed from the measurements that the time sampling for data collection is not constant which is due to limitation of smartphone sensors. Interpolation is one possible solution to address the issue of non-uniform sampling rate, that we also considered throughout this paper. 
	%In the experimental test, 40 participants were asked to work normally with their cell phones. They were asked to work with the cell phone in both portrait and landscape modes for one minute. The orientation data is measured for both sitting and mobile users. Here, is a summary of the experimental setup:
	
	\begin{itemize}
		\item Activities while sitting:
		\begin{enumerate}
			\item Browsing twice in portrait mode,
			\item Watching streaming videos twice in landscape mode,
		\end{enumerate}
		\item Activities while walking following a certain path:
		\begin{enumerate}
			\item Browsing in portrait mode,
			\item Watching streaming videos in landscape mode.
		\end{enumerate}
	\end{itemize}
	\noindent The path that the participants took was a straight corridor with dimensions of $40$ m $\times 1.5$ m. The participants were asked to walk down the corridor once. We note that the shape of the test area should not affect the experimental results and the model for the elevation angle as it mostly depends on the posture and physical attributes of typical users rather than the environment. This has been confirmed with sets of uncontrolled data collections from participants using their device in different environments \cite{ZhihongOrientation}. %The experimental results of this paper are mostly based on the controlled data set. However, we have also observed that the main conclusions of the paper are validated by uncontrolled experimental measurements as reported in \cite{ZhihongOrientation}.

	\begin{figure}
		\centering
		\begin{subfigure}[b]{0.5\columnwidth}
			\centering
			\includegraphics[width=1\columnwidth,draft=false]{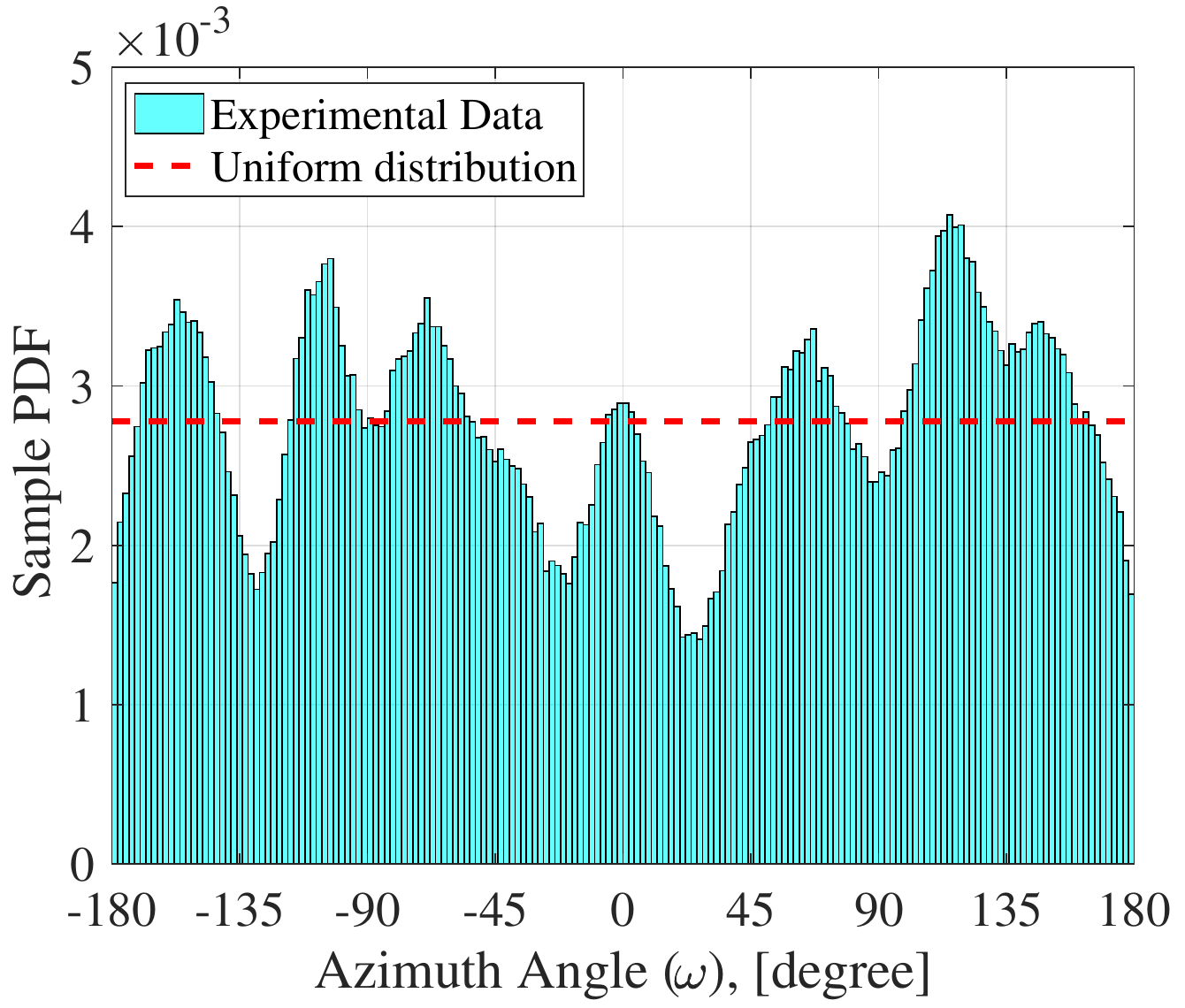}
			\caption{Sitting activities}
		\end{subfigure}%
		~
		\begin{subfigure}[b]{0.5\columnwidth}
			\centering
			\includegraphics[width=1\columnwidth,draft=false]{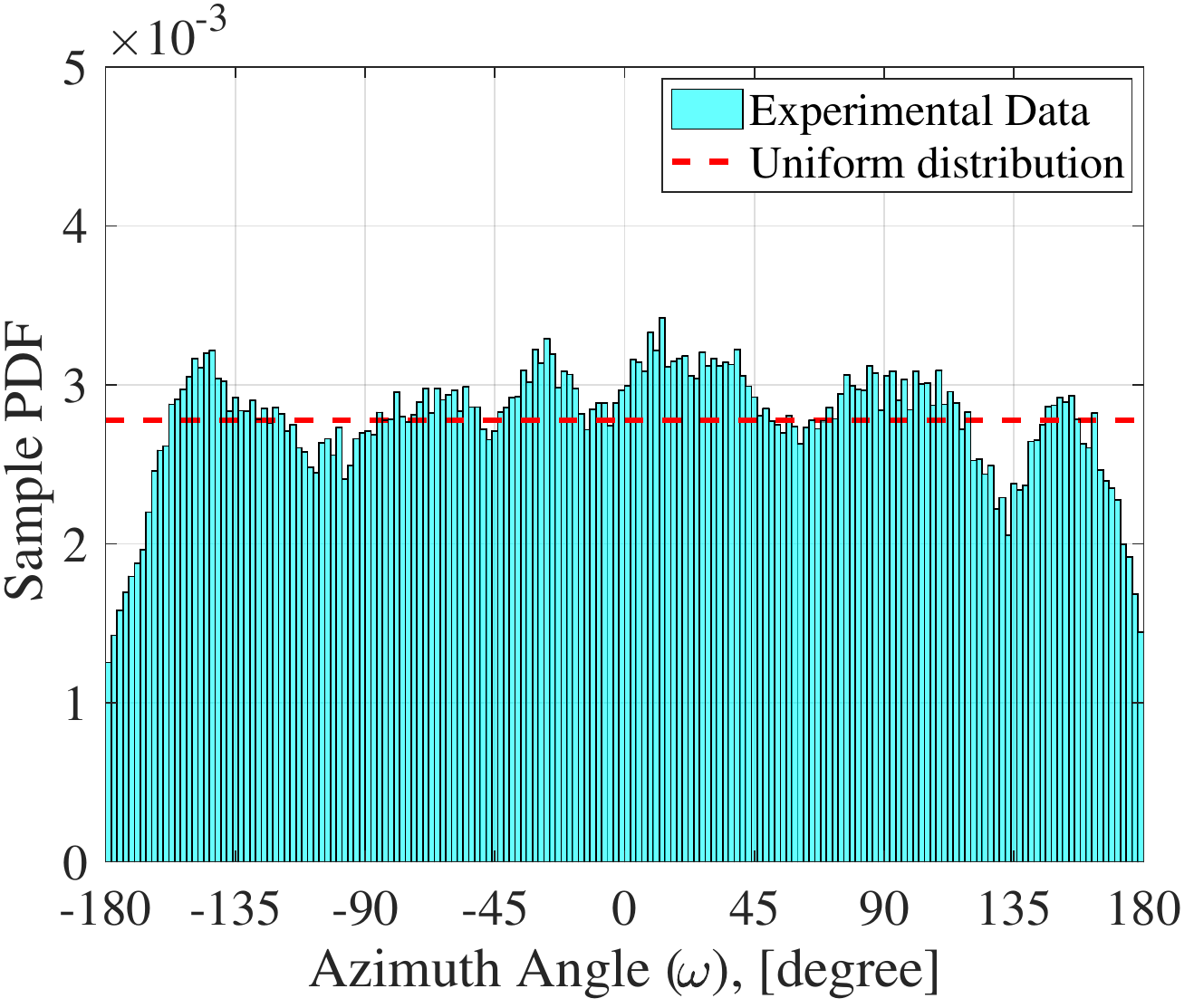}
			\caption{Walking activities}
		\end{subfigure}\\
		\begin{subfigure}[b]{0.5\columnwidth}
			\centering
			\includegraphics[width=1\columnwidth,draft=false]{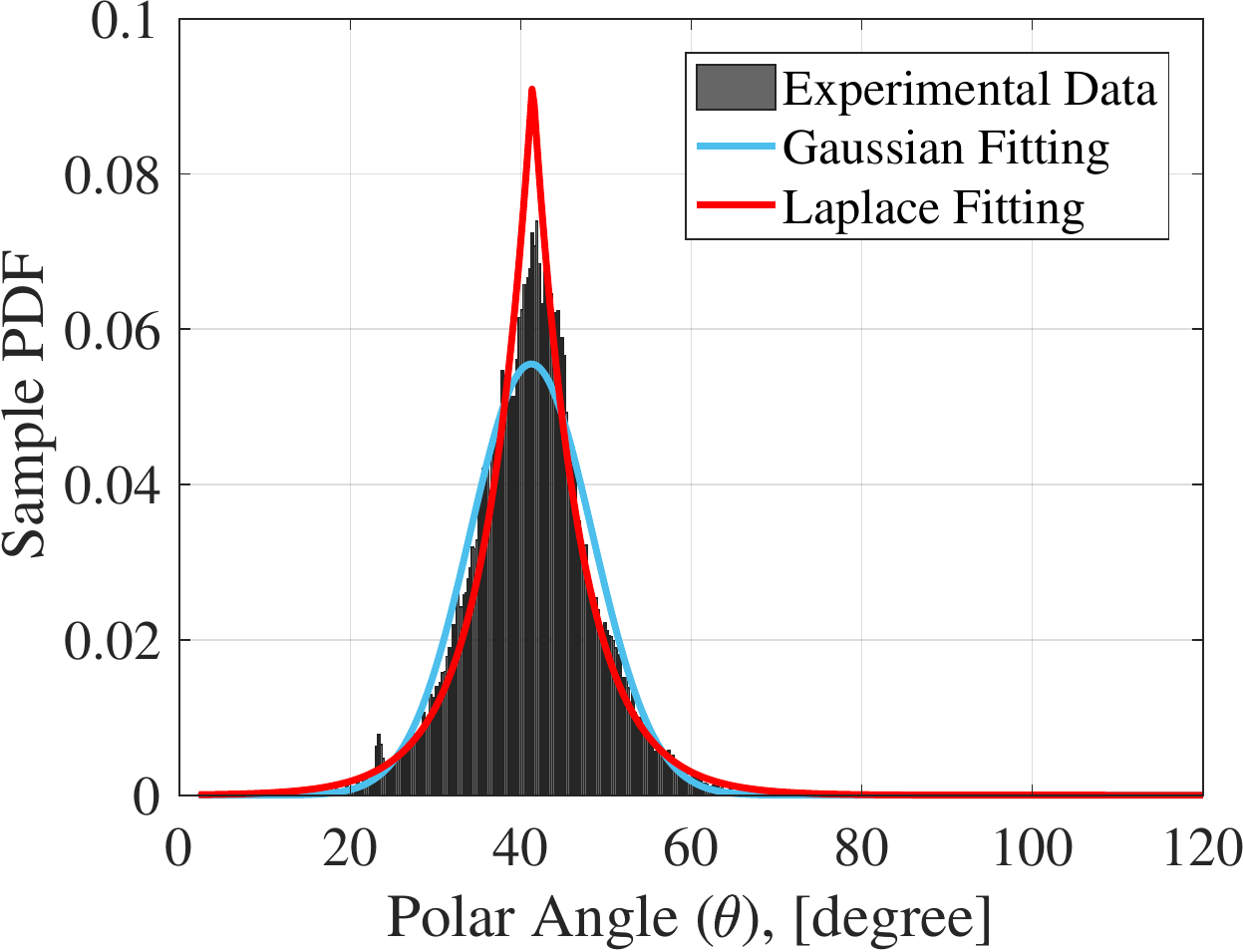}
			\caption{Sitting activities}
		\end{subfigure}%
		~
		\begin{subfigure}[b]{0.5\columnwidth}
			\centering
			\includegraphics[width=1\columnwidth,draft=false]{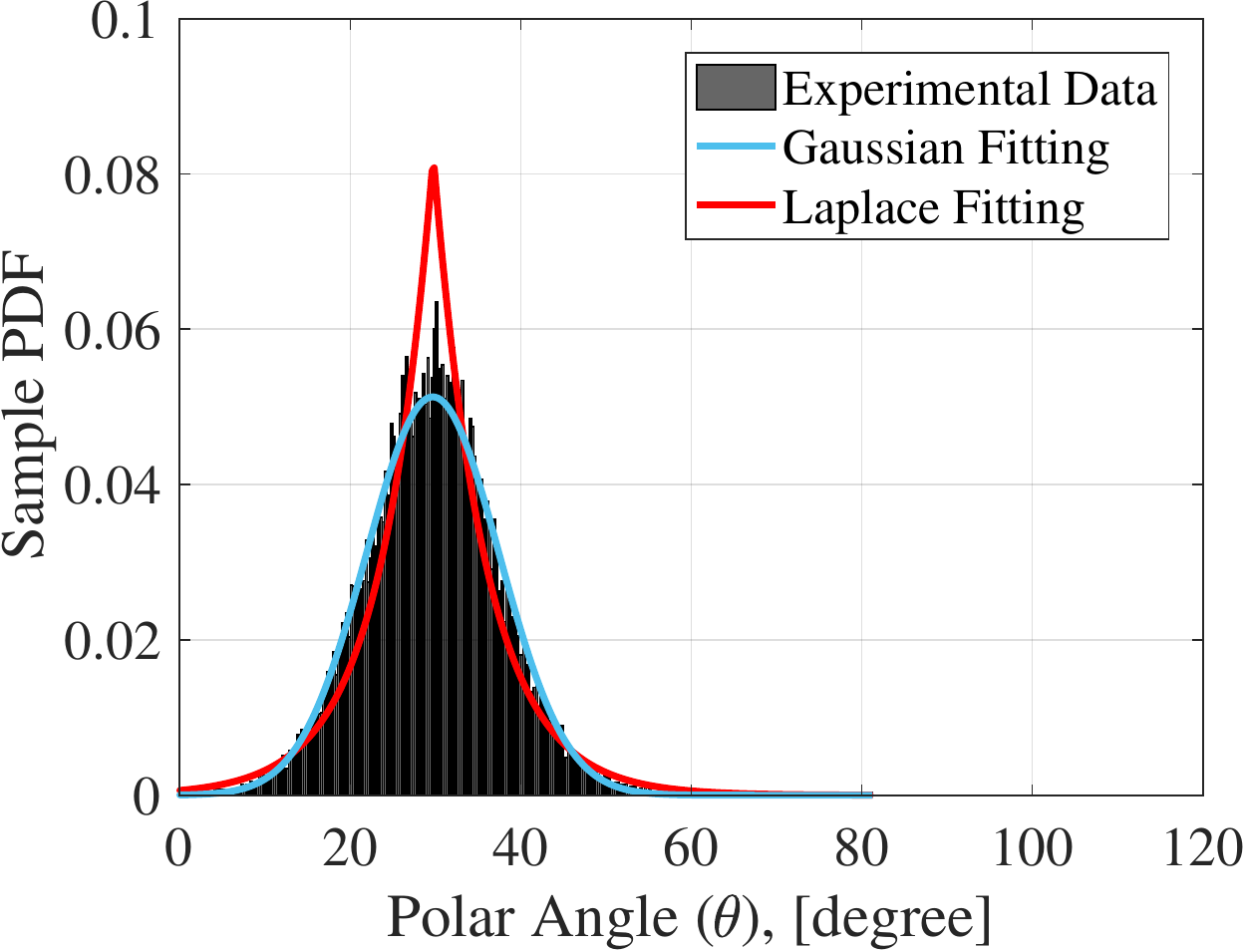}
			\caption{Walking activities}
		\end{subfigure}
		\caption{Samples PDFs of azimuth and polar angles with their distribution fitting. }
		\label{figsamplepdf}
	\end{figure}

	\vspace{-5pt}
	\subsection*{Measurement and Analysis}
	The statistics of the azimuth and polar angles can be measured based on the collected experimental data set. In order to evaluate the similarity of measurement data with a particular distribution, we consider Kolmogorov-Smirnov distance (KSD), skewness and kurtosis. The two-sample Kolmogorov-Smirnov distance (KSD) is the maximum absolute distance between the CDFs of two data vectors, which can be obtained as \cite{massey1951kolmogorov}, 
	\begin{equation}
	D=\max_x \left( \left|\hat{F}_1(x)-\hat{F}_2(x) \right| \right),
	\end{equation}
	where $\hat{F}_1(x)$ and $\hat{F}_2(x)$ are the CDFs of the first and second data vectors, respectively. Smaller values of KSD correspond to more similarity of the distributions.
	Skewness is described as a measure of the symmetry or asymmetry of a probability distribution. A perfectly symmetrical distribution will have a skewness of $0$. For example, the Gaussian and Laplace distributions both have a skewness of $0$. Mathematically, the skewness of a RV $X$ is defined as \cite{murray1972theory}:
	\begin{equation}
	{\rm Skew}[X]=\frac{\mathbb{E}\left[(x-\mu)^3\right] }{\sigma^3},
	\end{equation}
	where $\mu$ and $\sigma$ are the mean and variance of the distribution $X$. 
	Kurtosis is a measure of the tailedness of a probability distribution which is given as:
	\begin{equation}
	{\rm Kurt}[X]=\frac{\mathbb{E}\left[(x-\mu)^4 \right] }{\sigma^4}.
	\end{equation}
	The kurtosis of Laplace, Gaussian and Uniform distributions are $6$, $3$ and $1.8$, respectively. 
	
	The sample PDF of the azimuth angle for sitting and walking activities are represented in Fig.~\ref{figsamplepdf}-(a) and Fig.~\ref{figsamplepdf}-(b), respectively.  
	As can be seen, the azimuth angle closely follows a uniform distribution, i.e., $\omega\sim\mathcal{U}[-\pi,\pi)$, with the skewness of $-0.03$, kurtosis of $1.68$ and Kolmogorov-Smirnov distance (KSD) of $0.034$ for sitting activities. Also it is shown that walking activities have a skewness of $-0.0045$, kurtosis of $1.85$ and KSD of $0.019$. Note that the peaks in the PDFs (more visible for sitting) are due to the limited number of users, therefore, the azimuth angle is quite correlated for each sitting user along with the direction of the chair they used. 
	
	For the polar angle, $\theta$, we model it with Laplace and Gaussian distributions taking into account the first and second laws of error. According to the first law of error proposed by Laplace in $1774$, the frequency of an error can be represented as an exponential function of the numerical magnitude of the error, regardless of the sign as $f_X(x)=\frac{k}{2}e^{-k|x-x_q|}$, where $x$ and $x_q$ are the measured data and actual value, respectively and $k$ is a constant \cite{kotz2012laplace}. The second law of error was also proposed by Laplace  four years later and it states that the frequency of error is an exponential function of the square of the error, $f_X(x)=\frac{1}{\sqrt{2\pi}\sigma}e^{-\frac{(x-x_q)^2}{2\sigma^2}}$ with $\sigma$ as the variance of measured data. This is also called the normal distribution or Gauss law of error \cite{kotz2012laplace}. 		
%	These two laws of error satisfy the condition that positive and negative errors of equal magnitude are equally likely. In the following, we will consider both laws of error to model the polar angle, $\theta$. 
	
	The PDF of the polar angle, $\theta$, is estimated based on a set of almost uncorrelated data samples taken from the measurement data. Since the acquired data from the application are unevenly-spaced in time, we first generate a set of sufficiently separated data in time to ensure that samples taken from an individual user are almost  uncorrelated.  To define the proper separation between uncorrelated samples, we need to calculate the coherence lag between the samples that can be acquired by the autocorrelation function (ACF). However, due to uneven distribution of samples in time, the classical Fourier analysis is no longer valid. In fact, interpolation can be used to generate evenly spaced data samples but this would affect the autocorrelation by removing higher frequencies \cite{vanivcek1969approximate,baisch1999spectral}. Another solution would be the use of least-squares spectral analysis (LSSA) \cite{lomb1976least,vanderplas2017understanding}. Based on the LSSA method, we first calculate the power spectral density (PSD) of the unevenly spaced samples and then, the ACF can be estimated by calculating the inverse Fourier transform of the PSD. This technique has been explained in detail in our recent paper \cite{ArdimasOFDM} that reports the coherence time of the polar angle for sitting and walking activities as $373$ ms and $134$ ms, respectively.   
	
	The histogram of $\theta$ using \eqref{eqtheta}, where $\beta$ and $\gamma$ are collected from the experimental measurements is shown in Fig.~\ref{figsamplepdf}-(c) for sitting activities and in Fig.~\ref{figsamplepdf}-(d) for walking activities. The distribution fitting is performed based on the maximum likelihood estimator (MLE) of the Gaussian and Laplace distributions. The KSD is used to measure the distance between the Gaussian or Laplace distributions and the collection of datasets. The skewness and kurtosis of the collection of datasets are also calculated. 
	From Table~\ref{tabtheta}, the kurtosis of the polar angle, $\theta$, for sitting activities is $6.36$ which is closer to the kurtosis of the Laplace distribution than the Gaussian distribution, whereas for walking activities the kurtosis of empirical data is $3.77$ which confirms a greater similarity to the Gaussian distribution compared to the Laplace distribution.
	Hence, for sitting activities, the Laplace distribution closely matches the distribution of the experimental measurements in comparison with the Gaussian distribution when considering both KSD and kurtosis metrics. For walking activities, however, the Gaussian distribution matches the experimental data more closely. For the rest of the paper up to section~\ref{ORWP}, we mainly focus on the sitting activities and therefore the derivations are based on the Laplace distribution for the polar angle. However, one can readily apply a similar methodology to obtain the analytical results for the Gaussian distribution. Note that, in section~\ref{ORWP}, we use the Gaussian model because the UE is assumed to be mobile.  	
	The other important observation here is in regards to the moments of the measured data. The mean of the sitting and walking activities are about $41^{\circ}$ and $30^{\circ}$, respectively and both with a standard deviation of less than $9^{\circ}$.
	
	\begin{table}[] \centering \caption{Statistical Model of the Orientation of
			Mobile Devices} \label{tabtheta} \scalebox{0.85}{
			\begin{tabular}{|c|r|r|r|r|r|r|r|r|} \hline \multirow{3}{*}{}
				& \multicolumn{8}{c|}{Polar Angle ($\theta$) [degree]}
				\\ \cline{2-9}
				& \multicolumn{2}{c|}{Empirical Data}                        &
				\multicolumn{3}{c|}{\begin{tabular}[c]{@{}c@{}}The Gaussian\\
						Fitting\end{tabular}} &
				\multicolumn{3}{c|}{\begin{tabular}[c]{@{}c@{}}The Laplacian\\
						Fitting\end{tabular}} \\ \cline{2-9}
				& \multicolumn{1}{l|}{Skewness} & Kurtosis              & $\mu_{\text{G}}$
				& $\sigma_{\text{G}}$             & KSD            & $\mu_{\text{L}}$
				& $\sigma_{\text{L}}$            & KSD             \\ \hline
				\begin{tabular}[c]{@{}c@{}}Sitting\end{tabular}
				& $0.21                       $ & $6.36               $ & $41.23
				$ & $7.18                         $ & $0.04        $ & $41.39
				$ & $7.68                        $ & $0.04        $  \\ \hline
				\begin{tabular}[c]{@{}c@{}}Walking\end{tabular}
				& $0.13                       $ & $3.77               $ & $29.67
				$ & $7.78                         $ & $0.02        $ & $29.74
				$ & $8.59                        $ & $0.05        $  \\ \hline \end{tabular} }
		\vspace{-10pt}
	\end{table}

	Based on the experimental results, the PDF of $f_\theta\left(\theta \right)$ can be properly fitted with the truncated Laplace distribution. Mathematically, $\theta\in[0,\pi]$, however, as shown by the experimental measurements, the samples of the angle $\theta$ are restricted to the range $[0,\frac{\pi}{2}]$. Therefore, the PDF of $\theta$ can be conveniently denoted as:\vspace{0.0cm}
	\begin{align}
	\label{eqpdftheta}
	f_\theta(\theta) = \frac{\exp \left(-\frac{\left| \theta-\mu_{\theta}\right| }{b_{\rm{\theta}}}\right)}{2 b_{\rm{\theta}}\left(G(\frac{\pi}{2})-G(0) \right) }, \quad 0 \leq \theta \leq \frac{\pi}{2},
	\end{align}
	\noindent where $b_{\rm{\theta}} = \sqrt{\sigma^2_{\text{L}}/2} > 0$. The mean and scale parameters are set to the values from Table~\ref{tabtheta}. That is, $\mu_{\theta}=\locparam{L}$ and $b_{\rm{\theta}}=\scaleparam{L}$. Furthermore, $G(0)=\frac{1}{2}\exp\left(\frac{-\mu_{\theta}}{b_{\rm{\theta}}} \right)$ and $G(\frac{\pi}{2})=1-\frac{1}{2}\exp\left(-\frac{\frac{\pi}{2}-\mu_{\theta}}{b_{\rm{\theta}}} \right)$. Note that with the parameters for $\mu_{\theta}$ and $b_{\rm{\theta}}$ given in Table~\ref{tabtheta}, we have $G(\frac{\pi}{2})\approx1$ and $G(0)\approx0$. Thus, \eqref{eqpdftheta} can be simplified as: 
	\begin{equation}
	\label{eqpdfthetaApp}
	\tilde{f}_\theta(\theta) \cong \frac{\exp \left(-\frac{\left| \theta-\mu_{\theta}\right| }{b_{\rm{\theta}}}\right)}{2 b_{\rm{\theta}} }, \quad 0 \leq \theta \leq \frac{\pi}{2},
	\end{equation}
	where $\int_{0}^{\frac{\pi}{2}} \tilde{f}_\theta(\theta) ~\text{d}\theta \approx 1$. The CDF of $\theta$ is also given as:
	\begin{equation}
	\label{eqcdftheta}
	\tilde{F}_{\theta}(\theta)\cong\begin{cases}
	\frac{1}{2}\exp\left(\frac{\theta-\mu_{\theta}}{b_{\rm{\theta}}} \right),  \ \ \ \ \ \ \ \ \ \ \ \ \ \ \ \  \theta<\mu_{\theta} \\
	1-\frac{1}{2}\exp\left(-\frac{\theta-\mu_{\theta}}{b_{\rm{\theta}}} \right),  \ \ \ \ \ \ \ \ \ \  \theta\geq\mu_{\theta}
	\end{cases}.
	\end{equation}

	% \begin{figure}
	%     \begin{center}
	%         \includegraphics[width=\widthoffigs\columnwidth,draft=false]{./figs/figthetasitting_final.eps}
	%         \caption{PDF of polar angle ($\theta$) in the sitting activity.}
	%         \label{figthetasitting_final}
	%     \end{center}
	% \end{figure}
	
	% \begin{figure}
	%     \begin{center}
	%         \includegraphics[width=\widthoffigs\columnwidth,draft=false]{./figs/figthetawalking_final.eps}
	%         \caption{PDF of polar angle ($\theta$) in the walking activity.}
	%         \label{figthetawalking_final}
	%     \end{center}
	% \end{figure}

	% %%%%%%%%%%%%%%%%%%%%%%%%%%%%%%%%%%%%%%%%%%%%%%%%%%%%%%%%%%%%%%%%%%%%%%%
	% %%%%%%%%%%%%%%%%%%%%%%%%%%%%%%%%%%%%%%%%%%%%%%%%%%%%%%%%%%%%%%%%%%%%%%%
	% %%%%%%%%%%%%%%%%%%%%%%%%%%%%%%%%%%%%%%%%%%%%%%%%%%%%%%%%%%%%%%%%%%%%%%%
	%\section{Distribution of $\ccos{\psi}$}
	\begin{figure}
		\begin{center}
			\includegraphics[width=1\columnwidth,draft=false]{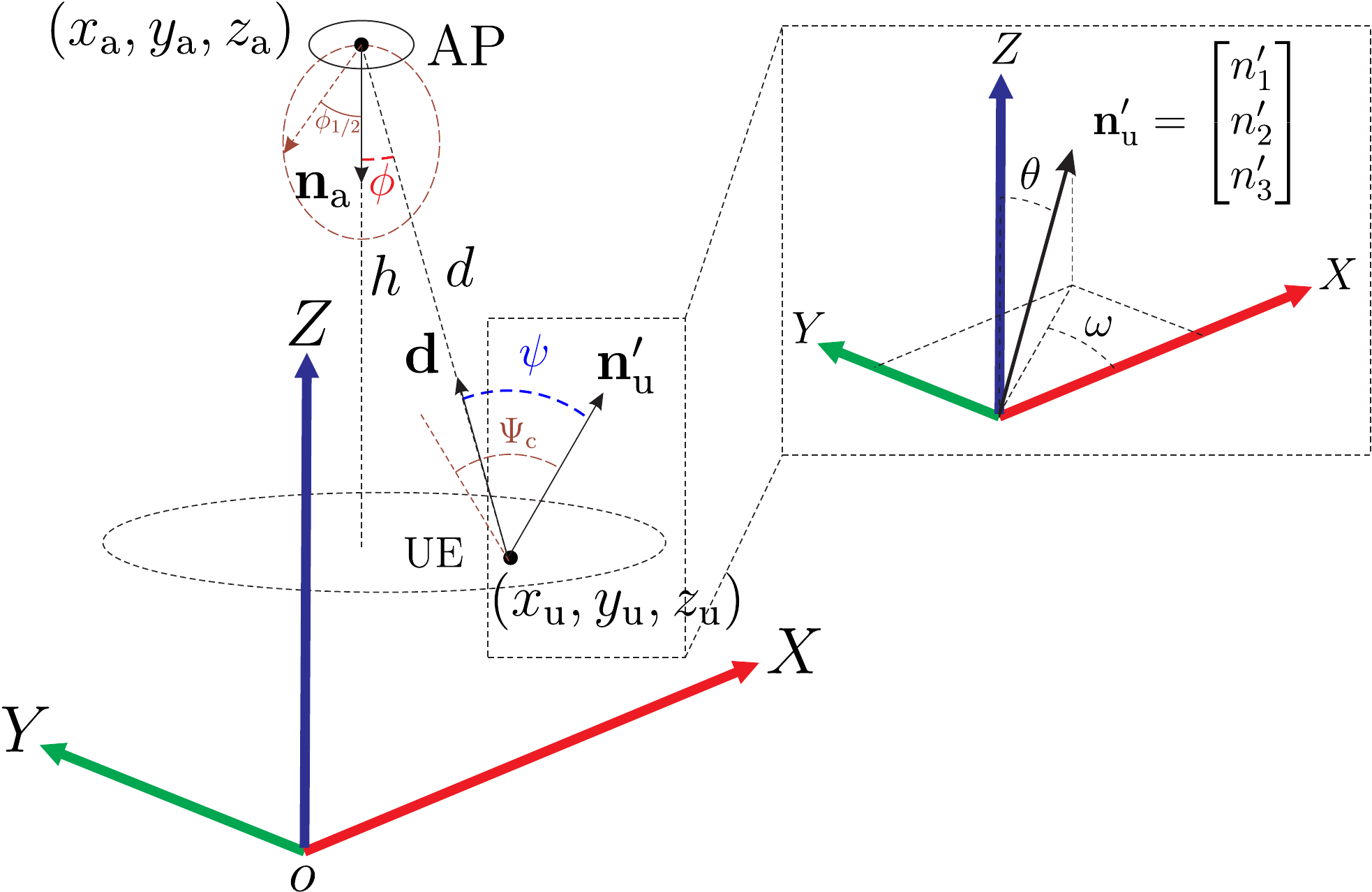}
			\caption{Downlink geometry of optical wireless communications with randomly-orientated UE and the spherical coordinates.}
			\vspace{-0.2cm}
			\label{figsphericalcoordinate}
		\end{center}
		\vspace{-15pt}
	\end{figure}
	
	\section{Analysis of Random Orientation Effect on Channel Gain}
	The downlink geometry of a generic optical wireless communication system is shown in Fig.~\ref{figsphericalcoordinate}. The locations of a UE and an AP are denoted by position vectors $\left( x_{\rm{u}}, y_{\rm{u}}, z_{\rm{u}} \right)$ and $\left( x_{\rm{a}}, y_{\rm{a}}, z_{\rm{a}} \right)$, respectively. The vertical distance of the AP and the UE is denoted by $h$, where $h = z_{\rm{a}} - z_{\rm{u}}$, and the Euclidean distance of the AP and the UE is denoted by $d$. The half-power semiangle of the LED is denoted as $\phi_{1/2}$ and the field-of-view (FOV) of the PD is $\Psi_{\rm{c}}$. The angle of incidence at the PD is denoted by $\psi$. 
	The LOS channel gain of the optical wireless channel is given as \cite{infraredkahnbarry}: 
	\begin{align}\label{eqpr}
	H = \frac{(m+1)A}{2\pi d^2} \cos^m\phi \cos\psi\ {\rm{rect}\left( \frac{\psi}{\Psi_{\rm c}}\right) } , 
	\end{align}
	where $m\!=\!\!-\ln(2)/\!\ln\left(\cos\!\left(\phi_{1/2}\right)\right)$, ${\rm{rect}(\frac{\psi}{\Psi_{\rm c}})}=1$ for $0\leq\! \psi \!\leq\! \Psi_{\rm c}$ and $0$ otherwise; $A$ is the physical area of the PD. From Fig.~\ref{figsphericalcoordinate} and \eqref{eqpr}, it is clear that for a particular UE position, the statistics of the LOS highly depend on the device orientation. In order to develop the statistics of the LOS channel gain, the statistics of $\cos{\psi}$ needs to be first determined based on the statistics of the device orientation discussed in section~\ref{ExperimentalSetup}. We note that the radiance angle, $\phi$, is not affected by the random orientation and we always have $\cos\phi=-\mathbf{n}_{\rm{a}} \cdot \mathbf{d}$. It should be mentioned that in this paper, we focus on the effect of UE's random orientation and do not consider the possible movements in the user's hands that may change the distance $d$. Note that such random changes in the distance would be small compared to the average distance between the AP and the UE.	Nevertheless, if an analysis of the random movement of the UE is of interest, our results can be readily used as the conditional statistics of the channel gain for a given location of the UE to develop the joint statistics assuming that the statistics of the random movement is available. This is similar to how we analyze the effect of random orientation in a mobile scenario in section~\ref{ORWP} where the random movement is modeled based on the random waypoint model.
	\vspace{-10pt}
	\subsection{PDF of $\cos{\psi}$}
	Throughout this paper, we calculate the PDF of $\cos{\psi}$ conditioned on $\omega$. 
	Let the vector $\mathbf{d}$ be the distance vector from the UE to the AP as shown in Fig.~\ref{figsphericalcoordinate}. Noting that the vector $\mathbf{n}'_{\rm{u}}$ denotes the normal vector of the UE receiver after rotation, the cosine of the incidence angle, $\psi$, can be obtained as:
	\begin{align}\label{eqcosphi}
	&\cos\psi = \mathbf{n}_{\rm{u}} \cdot \mathbf{d} = \left( \frac{x_{\rm{a}}-x_{\rm{u}}}{d} \right) \sin \theta \cos \omega + \nonumber\\
	&\left( \frac{y_{\rm{a}}-y_{\rm{u}}}{d} \right) \sin \theta \sin \omega+ \left(\frac{z_{\rm{a}}-z_{\rm{u}}}{d}\right) \cos \theta.
	\end{align}	
	
	\noindent where $\omega$ is given in \eqref{omegaeq} and depends on $\alpha$, $\beta$ and $\gamma$. According to the experimental data, the standard deviation of $\gamma$ in portrait mode for sitting and walking activities are $0.072$ and $0.113$ radian, respectively. The standard deviation of $\beta$ in landscape mode for sitting and walking activities are $0.044$ and $0.091$ radian, respectively. These small values of standard deviation confirm that when a user is working with the cellphone in portrait mode, the roll angle, $\gamma$, is almost zero. Whereas, when the user works with the cellphone in landscape mode, the pitch angle, $\beta$, is close to zero. 
	%However, one interesting point observed from experimental measurements is that when a user is working with the cellphone in portrait mode, the roll angle, $\gamma$, is close to zero. Further, when the user works with the cellphone in the landscape mode, the pitch angle, $\beta$, is close to zero. 
	Hence, substituting $\gamma=0$ in \eqref{omegaeq}, it can be simplified as $\hat{\omega}=\tan^{-1}\left(\frac{-\cos\alpha}{\sin\alpha} \right)$. Therefore, for portrait mode, we have:
	\begin{equation}
	\hat{\omega}=\begin{cases}
	\alpha-\frac{\pi}{2} \ \ \ \ \ \ \ \ \ \ {\alpha\in(0,\frac{3\pi}{2}]} \\
	\alpha-\frac{5\pi}{2} \ \ \ \ \ \ \ \ \  {\alpha\in(\frac{3\pi}{2},2\pi]}
	\end{cases}.
	\end{equation}
	Similarly for landscape mode, we have:
	\begin{equation}
	\hat{\omega}=\begin{cases}
	\alpha \ \ \ \ \ \ \ \ \ \ \ \ \ \ \ \ \ \ \ \ {\alpha\in(0,\pi]} \\
	\alpha-2\pi \ \ \ \ \ \ \ \ \ \ \ \ \ \  {\alpha\in(\pi,2\pi]}
	\end{cases}.
	\end{equation}
	Comparing $\Omega$ given in \eqref{OmegaPortrait} and \eqref{OmegaLandscape}, with the above equations for $\hat{\omega}$, one can easily deduce $\Omega=\hat{\omega}+\pi$. Note that, based on the measurements, the mean absolute error of approximating  the angles $\omega$ as $\hat{\omega}$ is about 0.09 and 0.14 radian respectively for sitting and mobile users, which confirms a relatively good accuracy of the approximation. For the rest of the paper, we consider the angle $\Omega$ in the equations since it has a better physical interpretation compared to the angles $\hat{\omega}$ or $\omega$. As noted before, the angle $\Omega$ represents the angle of the direction the user is facing and the $X$-axis of Earth coordinates. Therefore, \eqref{eqcosphi} can be approximated as $\cos\psi\cong -\left( \frac{x_{\rm{a}}-x_{\rm{u}}}{d} \right) \sin \theta \cos \Omega -\left( \frac{y_{\rm{a}}-y_{\rm{u}}}{d} \right) \sin \theta \sin \Omega+ \left(\frac{z_{\rm{a}}-z_{\rm{u}}}{d}\right) \cos \theta$. For simplicity of notation, this equation can be rewritten as:
	\begin{align}\label{eqcosphigivenomega}
	g(\theta) &\triangleq \cos{\psi} = a \sin{\theta} + b \cos{\theta},
	\end{align}
	\noindent where
	\begin{align*}
	a &= -\left( \frac{x_{\text{a}}-x_{\rm{u}}}{d} \right) \cos{\Omega} - \left( \frac{y_{\text{a}}-y_{\rm{u}}}{d} \right) \sin{\Omega},\  \text{and}\ b = \left(\frac{z_{\text{a}}-z_{\rm{u}}}{d}\right).
	\end{align*}	
	
	The coefficients $a$ and $b$ play a prominent role in the analysis of $\cos\psi$ so it is worth investigating them in detail to help readers intuitively understand them.  
	Throughout this paper, the AP is always located above the UE, i.e., $b > 0$. The coefficient $a$ can be rewritten as: 
	\begingroup\makeatletter\def\f@size{9.3}\check@mathfonts
	\def\maketag@@@#1{\hbox{\m@th\large\normalfont#1}}%
	\begin{align*}
	a = \sqrt{\frac{\left(x_{\text{a}}-x_{\rm{u}}\right)^2 + \left(y_{\text{a}}-y_{\rm{u}}\right)^2}{d^2}} \cos \left( \Omega - \tan^{-1}\left(\frac{y_{\text{u}}-y_{\rm{a}}}{x_{\text{u}}-x_{\rm{a}}}\right) \right),
	\end{align*} \endgroup
	\noindent Fixing the AP location, the peak of the coefficient $a$ only depends on the position of the UE. Meanwhile, the sign of coefficient $a$ is negative if the UE does not face the AP (opposite direction to the AP). For example, if the AP is located in the position $(0,0,2)$, and the UE is located in $(-1,0,0)$, then $a$ will be negative if $\frac{-\pi}{2} < \Omega < \frac{\pi}{2}$. In addition, the coefficients $a$ and $b$ should satisfy the following inequalities:
	\begin{align}\label{eqineqab}
	-1 < a < 1,~ 0 < b \leq 1, ~\text{and}~ b \leq \sqrt{a^2+b^2} \leq 1.
	\end{align}
	
	Based on the sign of $a$, the term $\cos{\psi}$ may be a monotonically decreasing function of $\theta$ (\textbf{Case 1}) or it may be a concave downward function with one peak (\textbf{Case 2}). These two cases are explained next and the PDF of $\cos{\psi}$ is derived for each case. The detailed derivations are provided in Appendix\ref{App-A}.
	
	\noindent1) \textit{For $a < 0$ (\textbf{Case 1})}:
	If $a < 0$, $b > 0$ and $0 < \theta < \frac{\pi}{2}$, it can be seen from \eqref{eqcosphigivenomega} that $\cos{\psi}$ is a monotonically decreasing function of $\theta$. Using the fundamental theorem to calculate the PDF of a function of an RV given in \cite{papoulis1985random}, we get the PDF of $\cos{\psi}$ as:
	\begin{align}\label{eqfcosphigivenomegacase1}
	f_{\cos{\psi}}(\tau) = \frac{f_{\theta}\left( -\sin^{-1}\left( \frac{\tau}{\sqrt{a^2 + b^2}} \right)   -\tan^{-1}\left( \frac{b}{a} \right) \right)}{\sqrt{a^2+b^2-\tau^2}}, ~ a < \tau < b, 
	\end{align}
	\noindent where $\tau$ denotes the realization of the RV $\cos{\psi}$. Let's define: 
	\begin{align}
	\label{supremum}
	\text{ss}_{ f } \triangleq \sup \text{supp}\left( f_{\cos{\psi}} \right)
	\end{align}
	\noindent as the supremum of the support of $f_{\cos{\psi}}$. This metric is presented to emphasize that $\text{ss}_{ f }$ is not always $1$. For $a < 0$, $\text{ss}_{ f } = b$ which is strictly less than $1$.
	
	\noindent2) \textit{For $a \geq 0$ (\textbf{Case 2})}:
	If $a \geq 0$, $b > 0$, and $0 < \theta < \frac{\pi}{2}$, (\ref{eqcosphigivenomega}) is a concave downward function. Then, the maximum value of (\ref{eqcosphigivenomega}) is at the point $\theta^*$ given as follows:
	\begin{align}\label{eqthetastar}
	\theta^* = \argmax_{0\ <\ \theta \ <\ \frac{\pi}{2}} \left( a \sin{\theta} + b \cos{\theta}\right)  = \tan^{-1}\left( \frac{a}{b} \right). 
	\end{align}
	\noindent Therefore, the PDF of $\cos{\psi}$ can be expressed as in (\ref{eqfcosphigivenomegacase2}) for $g(0)\leq g(\frac{\pi}{2})$ or in (\ref{eqfcosphigivenomegacase3}) for $g(\frac{\pi}{2})< g(0)$, which are given at the top of the next page. Note that for $a \geq 0$, $\text{ss}_{ f } = \sqrt{a^2 + b^2} \leq 1$.

	\begin{floatEq}
		\begin{subequations}
			\label{eqfcosphigivenomegacase}
			\begin{equation}
			\label{eqfcosphigivenomegacase2}
			f_{\cos{\psi}}(\tau)=
			\begin{cases}
			\frac{f_{\theta}\left( \sin^{-1}\left( \frac{\tau}{\sqrt{a^2 + b^2}} \right) -\tan^{-1}\left( \frac{b}{a} \right)  \right)}{\sqrt{a^2+b^2-\tau^2}}, ~&g(0) < \tau < g(\frac{\pi}{2}) \\
			\frac{f_{\theta}\left( \sin^{-1}\left( \frac{\tau}{\sqrt{a^2 + b^2}} \right) -\tan^{-1}\left( \frac{b}{a} \right)  \right)}{\sqrt{a^2+b^2-\tau^2}} + \frac{f_{\theta}\left( -\sin^{-1}\left( \frac{\tau}{\sqrt{a^2 + b^2}} \right) -\tan^{-1}\left( \frac{b}{a} \right) + \pi \right)}{\sqrt{a^2+b^2-\tau^2}}, ~&g(\frac{\pi}{2}) < \tau \leq g(\theta^*)
			\end{cases}
			\end{equation}
			\begin{equation}
			\label{eqfcosphigivenomegacase3}
			f_{\cos{\psi}}(\tau)=\begin{cases}
			\frac{f_{\theta}\left( -\sin^{-1}\left( \frac{\tau}{\sqrt{a^2 + b^2}} \right) -\tan^{-1}\left( \frac{b}{a} \right) + \pi \right)}{\sqrt{a^2+b^2-\tau^2}}, ~&g(\frac{\pi}{2}) < \tau < g(0) \\
			\frac{f_{\theta}\left( \sin^{-1}\left( \frac{\tau}{\sqrt{a^2 + b^2}} \right) -\tan^{-1}\left( \frac{b}{a} \right)  \right)}{\sqrt{a^2+b^2-\tau^2}} + \frac{f_{\theta}\left( -\sin^{-1}\left( \frac{\tau}{\sqrt{a^2 + b^2}} \right) -\tan^{-1}\left( \frac{b}{a} \right) + \pi \right)}{\sqrt{a^2+b^2-\tau^2}}, ~&g(0) < \tau \leq g(\theta^*)
			\end{cases}
			\end{equation}	
		\end{subequations}
		\vspace{-15pt}	
	\end{floatEq}

	% \begin{align}\label{eqfcosphigivenomegacase2}
	% f_{\ccos{\psi}}(\tau)=
	%     \begin{cases}
	%         \frac{f_{\theta}\left( \sin^{-1}\left( \frac{\tau}{\sqrt{a^2 + b^2}} \right) -\tan^{-1}\left( \frac{b}{a} \right)  \right)}{\sqrt{a^2+b^2-\tau^2}}, ~g(0,a,b) < \tau < g(\theta^*,a,b) \\
	%         \frac{f_{\theta}\left( \sin^{-1}\left( \frac{\tau}{\sqrt{a^2 + b^2}} \right) -\tan^{-1}\left( \frac{b}{a} \right)  \right)}{\sqrt{a^2+b^2-\tau^2}} + \frac{f_{\theta}\left( -\sin^{-1}\left( \frac{\tau}{\sqrt{a^2 + b^2}} \right) -\tan^{-1}\left( \frac{b}{a} \right) + \pi \right)}{\sqrt{a^2+b^2-\tau^2}}, \\
	%         \quad\quad\quad\quad\quad\quad\quad\quad\quad~ g(\frac{\pi}{2},a,b) < \tau < g(\theta^*,a,b).
	%     \end{cases}
	% \end{align}
	
	It is worth noting that the PDF expressions given for the two cases do not particularly depend on whether $\theta$ is Laplace or Gaussian distributed so one can apply either of the distributions into \eqref{eqfcosphigivenomegacase1} and \eqref{eqfcosphigivenomegacase} and calculate the distribution of $\cos \psi$. %Fig.~\ref{figbehaviourpdf} presents the PDF of $\cos{\psi}$ given in (\ref{eqfcosphigivenomegacase1}) and (\ref{eqfcosphigivenomegacase2}) for	different positions of the UE with $\theta$ following the Laplace distribution. The UE is assumed to be stationary with $\Omega = \pi$ and located in the $XY$-plane, i.e., $z_{\text{u}} = 0$. The AP location is $(x_{\text{a}},y_{\text{a}},z_{\text{a}})=(0,0,2)$.	As can be seen from Fig.~\ref{figbehaviourpdf}, the analytical results match the experimental results. Taking a close look at the results, it can be inferred that for $a\leq0$,  $f_{\cos{\psi}}$ is still closely Laplace distributed. Although, for $a>0$ and especially when it is close to $1$, the shape of $f_{\cos{\psi}}$ does not resemble a Laplace distribution, we will show in the next section that it still can be approximated with a truncated Laplace distribution. 

	\subsection{Approximate PDF of $\cos\psi$}
	
	In order to make the performance analysis of OWC systems with random orientation tractable, one would be interested in approximating the closed-form PDF equations (\ref{eqfcosphigivenomegacase1}) and (\ref{eqfcosphigivenomegacase}) with simpler expressions. In this section, we show that the truncated Laplace distribution can be used to approximate the PDF of cosine of the incidence angle, $f_{\cos{\psi}}$ for the sitting scenario. Assuming a walking scenario, an approximation of the exact PDF as a truncated Gaussian distribution can be made similarly but was not included here.
	
	Note that the non-zero mean random variable $\theta$ can be expressed as $\theta=\mu_{\rm{L}}+\theta'$, where $\theta'$ follows a zero mean Laplace distribution with the same variance as $\theta$. Substituting $\theta=\mu_{\rm{L}}+\theta'$ in \eqref{eqcosphigivenomega}, and expanding the cosine and sine functions, we have:
	\begin{equation}
	\label{CosineExpanded}
	\begin{aligned}
	\cos\psi= ~&a\sin\theta'\cos\mu_{\theta}+a\sin\mu_{\theta}\cos\theta'+b\cos\theta'\cos\mu_{\theta}\\-
	&b\sin\theta'\sin\mu_{\theta}.
	\end{aligned}
	\end{equation}
	Under the condition of a small variance for $\theta$, and employing the first-order Taylor series approximation $(\sin\theta'\cong\theta'$ and $\cos\theta'\approx1)$,  \eqref{CosineExpanded} can be approximated as:
	\begin{equation}
	\label{CosineTaylor}
	\begin{aligned}
	&\!\!\!\!\!\!\!\!\cos\psi\cong
	a\theta'\cos\mu_{\theta}+a\sin\mu_{\theta}+b\cos\mu_{\theta}-b\theta'\sin\mu_{\theta}\\ &=\left(a\cos\mu_{\theta}\!-b\sin\mu_{\theta}\right)\theta'\!+a\sin\mu_{\theta}\!+\!b\cos\mu_{\theta}\!\triangleq\hat{g}(\theta').
	\end{aligned}
	\end{equation}
	Noting that $\hat{g}(\theta')$ is a linear function of $\theta'$ and $\hat{\tau}_{\rm{min}}\leq\cos\psi\leq \hat{\tau}_{\rm{max}}$, the PDF of cosine of the incidence angle can be expressed based on the truncated Laplace distribution given as follows:
	\begin{equation}
	\label{PDFCosineTaylor}
	\tilde{f}_{\hat{g}}(\hat{\tau})=\frac{f_{\theta'}(\hat{\tau})}{F_{\theta'}(\hat{\tau}_{\rm{max}})-F_{\theta'}(\hat{\tau}_{\rm{min}})},\ \ \  \hat{\tau}_{\rm{min}}\leq\hat{\tau}\leq \hat{\tau}_{\rm{max}},
	\end{equation}
	where $\hat{\tau}_{\rm{min}}=-1$ and taking into account the supremum of the support of $f_{\cos\psi}$, then, $\hat{\tau}_{\rm{max}}=b$ for $a<0$ and $\hat{\tau}_{\rm{max}}=\sqrt{a^2+b^2}$ for $a\geq0$. In \eqref{PDFCosineTaylor}, $\theta'$ follows a Laplace distribution with the following parameters:
	\begin{equation}
	\label{TaylorParametersMean}
	\hat{\mu_{\theta}} = a \sin{\mu_{\theta}} + b \cos{\mu_{\theta}}, 
	\end{equation}
	\begin{equation}
	\label{TaylorParametersb}
	\hat{b_{\rm{\theta}}} =  b_{\rm{\theta}} \left| a \cos{\mu_{\theta}} - b \sin{\mu_{\theta}} \right|.
	\end{equation}
	
		\begin{figure}[t!]
			\begin{center}
				\includegraphics[width=1\columnwidth,draft=false]{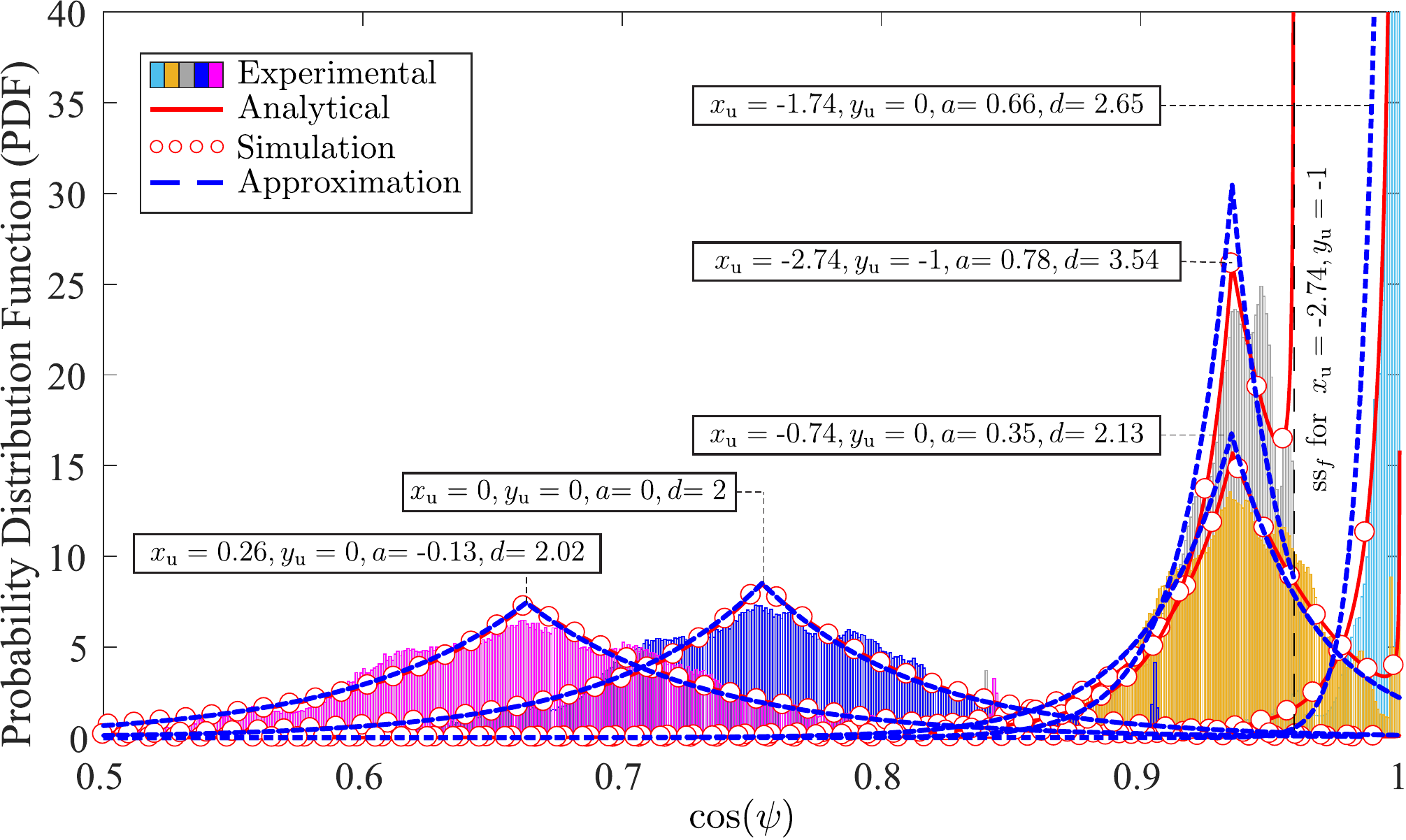}
				\vspace{-10pt}
				\caption{Behavior of the sample PDF of experimental data and the closed-form PDFs with different locations of UE and fixed $(x_{\text{a}},y_{\text{a}},z_{\text{a}})=(0,0,2)$, $z_{\text{u}} = 0$, $\Omega = \pi$, and for the sitting activity.}
				\label{figbehaviourpdf}
			\end{center}
			\vspace{-20pt}
		\end{figure}
	
	Substituting $\hat{\tau}_{\rm{min}}$ and $\hat{\tau}_{\rm{max}}$ in \eqref{PDFCosineTaylor}, the truncated Laplace distribution of $\cos\psi$ conditioned on $\Omega$ is given as:
	\begin{equation}
	\begin{aligned}
	\label{PDFCosineTaylorTrunc}
	\tilde{f}_{\cos\psi}(\hat{\tau})&=\frac{1}{\Delta(\hat{\mu_{\theta}},\hat{b_{\rm{\theta}}},\hat{\tau}_{\rm{max}})}\exp\left(-\frac{|\hat{\tau}-\hat{\mu_{\theta}}|}{\hat{b_{\rm{\theta}}}} \right)
	\end{aligned}
	\end{equation}
	where $\Delta(\hat{\mu_{\theta}},\hat{b_{\rm{\theta}}},\hat{\tau}_{\rm{max}})\!=\!2\hat{b_{\rm{\theta}}}\left(1\!-\frac{1}{2}\exp\left(\!\frac{\hat{\mu_{\theta}}-\hat{\tau}_{\rm{max}}}{\hat{b_{\rm{\theta}}}}\! \right)-\frac{1}{2}\exp\left(\!\frac{-1-\hat{\mu_{\theta}}}{\hat{b_{\rm{\theta}}}}\! \right) \right)$ and the support range of $\tilde{f}_{\cos\psi}$ is $-1\leq\hat{\tau}\leq \hat{\tau}_{\rm{max}}$. Furthermore, the corresponding CDF can be obtained by simplifying the $\tilde{F}_{\cos\psi}(\hat{\tau})=\int_{-1}^{\hat{\tau}}\tilde{f}_{\cos\psi}(x){\rm{d}}x$, as follows:
	\begin{equation}
	\tilde{F}_{\cos\psi}(\hat{\tau})\!=\!
	\begin{cases}
	\!\frac{\hat{b_{\rm{\theta}}}}{\Delta(\hat{\mu_{\theta}},\hat{b_{\rm{\theta}}},\hat{\tau}_{\rm{max}})}\left( \exp\left( \frac{\hat{\tau}-\hat{\mu_{\theta}}}{\hat{b_{\rm{\theta}}}} \right)-\exp\left( \frac{-1-\hat{\mu_{\theta}}}{\hat{b_{\rm{\theta}}}}\right)\right)  , & \hat{\tau}< \hat{\mu_{\theta}} \\
	\!\frac{\hat{b_{\rm{\theta}}}}{\Delta(\hat{\mu_{\theta}},\hat{b_{\rm{\theta}}},\hat{\tau}_{\rm{max}})}\!\left(2-\exp\left( \frac{\hat{\mu_{\theta}}-\hat{\tau}}{\hat{b_{\rm{\theta}}}} \right)-\exp\left( \frac{-1-\hat{\mu_{\theta}}}{\hat{b_{\rm{\theta}}}}\right) \right) , & \hat{\tau}\geq \hat{\mu_{\theta}}
	\end{cases}
	\end{equation}
	
	%Note that since $\hat{\tau}_{\rm{min}}<< \hat{\mu_{\theta}}$ for all $a$ and $b$, so $F_{\theta'}(\hat{\tau}_{\rm{min}})\cong 0$. Therefore, \eqref{PDFCosineTaylorTrunc} can be simplified further as:
	%\begin{equation}
	%\begin{aligned}
	%\label{PDFCosineTaylorTruncSimpl}
	%f_{\cos\psi}(\hat{\tau})=\frac{\exp\left(\frac{|\hat{\tau}-\hat{\mu_{\theta}}|}{\hat{b_{\rm{\theta}}}} \right) }{2\hat{b_{\rm{\theta}}}\left(1-\frac{1}{2}\exp\left(\frac{|\hat{\mu_{\theta}}-1|}{\hat{b_{\rm{\theta}}}} \right) \right) }, 
	%\end{aligned}
	%\end{equation} 

	In order to evaluate the accuracy of our proposed truncated Laplace model based on the first-order Taylor series, we have calculated the average KSD with respect to the experimental data over different positions in a room of size $10\times10$ m$^2$. The average KSD for sitting activities of the proposed model given in \eqref{PDFCosineTaylorTrunc} and the exact PDF given in \eqref{eqfcosphigivenomegacase1} and \eqref{eqfcosphigivenomegacase} are $0.055$ and $0.026$, respectively. 
	These values are relatively small so that our exact derived PDF and the proposed approximation based on the first order Taylor series show a close similarity with the sample PDF from the experimental measurements. Note that, as indicated before, an approximation of the exact PDF for the walking scenario as truncated Gaussian distribution can be made similarly using the linearity of $\hat{g}(\theta')$ in \eqref{CosineTaylor}.    

	Fig.~\ref{figbehaviourpdf} presents a comparison between the PDF of $\cos{\psi}$ given in (\ref{eqfcosphigivenomegacase1}) and (\ref{eqfcosphigivenomegacase}) with an approximate PDF expressed in \ref{PDFCosineTaylorTrunc} for
	different positions of the UE with $\theta$ following the Laplace distribution. The UE is assumed to be stationary with $\Omega = \pi$ and located in the $XY$-plane, i.e., $z_{\text{u}} = 0$. The AP location is $(x_{\text{a}},y_{\text{a}},z_{\text{a}})=(0,0,2)$.
	As can be seen, the first order Taylor approximate PDF provides a well-matched approximation with the experimental measurements. Taking a close look at the results, it can be inferred that for $a\leq0$,  $f_{\cos{\psi}}$ is still closely Laplace distributed. Although, for $a>0$ and especially when it is close to $1$, the shape of $f_{\cos{\psi}}$ does not resemble a Laplace distribution, however, it still can be approximated with a truncated Laplace distribution.

	\subsection{Behavior of the PDF of $\cos{\psi}$}
In this subsection, we give a deeper analysis on the behavior of the exact PDF of $\cos{\psi}$ to provide more justifications to support the approximation proposed in the previous subsection. Our analytical approach is to identify the critical points of $f_{\cos{\psi}}$ and analyze them. For both Laplace and Gaussian distributions, we expect a peak's location (referred as $\tau^*$) to strictly fall between the boundaries of the support of $f_{\cos{\psi}}$. To further justify the approximation of the exact PDF as either Laplace or Gaussian distributions, it will be shown that the rate of change of $f_{\cos{\psi}}$ is exponentially increasing and decreasing respectively for $\tau<\tau^*$ and $\tau>\tau^*$.
	
	We classify the analysis of the behavior of PDF of $\cos{\psi}$ into two discussions, i.e., for case 1 ($a < 0$) and for case 2 ($a \geq 0$). 
	
	\noindent1) \textit{For $a < 0$ (\textbf{Case 1})}:
	For case 1, we need to focus on \eqref{eqfcosphigivenomegacase1} whose support is the interval $(a,b)$. 
	
	\begin{proposition}[Proposition 1] 
		For $a < 0$, $f_{\cos{\psi}}$ has following characteristics: 
		\begin{enumerate}
			\item $\tau^*$ is the global maximum of $f_{\cos{\psi}}$ and always exists,  
			\item $f_{\cos{\psi}}$ is well defined for $a < \tau < b$,
			\item $f_{\cos{\psi}}(\tau)$ increases in an exponential function for $a < \tau < \tau^*$ and decreases in an exponential function for $\tau^* < \tau < b$, if $b_\theta < \min\left\{-b/a,-a/b \right\}$, and 
			\item $f_{\cos{\psi}}$ is continuous at $\tau = \tau^*$.
		\end{enumerate}
		\label{prop1}
	\end{proposition}
	
	\noindent A proof of Proposition 1 is provided in Appendix\ref{App-prop1}. To have a physical meaning of the conditions in Proposition 1 in terms of room dimension, they can be translated into following. In an indoor room with dimensions of $L \times L$ m$^2$,\footnote{These dimensions can also be interpreted as the domain of the locations of the UE.} $h = z_{\text{a}} - z_{\text{u}}$, and with the AP located in the center of the room, the following condition needs to be met:
	\begin{align}\label{eqineqblcase1}
	b_{\rm{\theta}} < \min\left\{ \frac{\sqrt{2} h}{L}, \frac{L}{\sqrt{2} h} \right\}.
	\end{align}
	\noindent To get the above condition, we use $\eqref{eqineqab}$ and the definition of $b$. For $L = 20$ m and $h = 2$ m, $b_{\rm{\theta}}$ should be less than $0.1414$ rad which is always true based on our experimental data. In addition, as the PDF $f_{\cos{\psi}}$ increases exponentially in the left tail, i.e., $a < \tau < \tau^*$, and decreases exponentially in the right tail, i.e., $\tau^* < \tau < b$, the truncated Laplace distribution can be used to approximate \eqref{eqfcosphigivenomegacase1}.
% * <majid.safari@ed.ac.uk> 2018-05-15T23:59:13.998Z:
% 
% > In addition, as the PDF $f_{\cos{\psi}}$ increases exponentially in the left tail, i.e., $a < \tau < \tau^*$, and decreases exponentially in the right tail, i.e., $\tau^* < \tau < b$, the truncated Laplace distribution can be used to approximate"
%  
% Can we say truncated Laplace or Gaussian distribution ...? I am not generally comfortable with the notion of exponential rate of change. What I understand from the term exponential rate is simply exp(-x) and for example not exp(-x^2) so I don't know how we cam combine Laplace and Gaussian here and more importantly so you really show exponential decay?  In the proof also you do not clearly discuss the exponential rate of decay. 
% 
% ^.

	\noindent2) \textit{For $a \geq 0$ (\textbf{Case 2})}:
	For case 2, we need to focus on \eqref{eqfcosphigivenomegacase} that its support is the interval $\left( \min\left\{g(0),g\left(\frac{\pi}{2}\right) \right\},g(\theta^*)\right)$. Note that $g(0) = b$, $g\left(\frac{\pi}{2}\right) = a$, and $g(\theta^*) = \sqrt{a^2+b^2}$.
	
	\begin{proposition}[Proposition 2] 
		For $a \geq 0$, $f_{\cos{\psi}}$ has following characteristics: 
		\begin{enumerate}
			\item $\tau^*$ does not always exist in the support $\left(\min\{a,b\},\sqrt{a^2+b^2}\right)$;  
			\item $f_{\cos{\psi}}$ is well defined for $\min\{a,b\} < \tau < \sqrt{a^2+b^2}$;
			\item when $\tau^*$ exists in the support $\left(\min\{a,b\},\sqrt{a^2+b^2}\right)$, $f_{\cos{\psi}}(\tau)$ increases in an exponential rate for $\min\{a,b\} < \tau < \tau^*$, decreases in an exponential rate for $\tau^* < \tau < \sqrt{\frac{a^2+b^2}{1+b_{\rm{\theta}}^2}}$, and increases in an exponential function for $ \sqrt{\frac{a^2+b^2}{1+b_{\rm{\theta}}^2}} < \tau < \sqrt{a^2+b^2}$; 
			\item when $\tau^*$ exists in the support $\left(\min\{a,b\},\sqrt{a^2+b^2}\right)$, $f_{\cos{\psi}}(\tau)$ increases in an exponential function for $\min\{a,b\} < \tau < \sqrt{a^2+b^2}$; and 
			\item $f_{\cos{\psi}}$ is continuous at $\tau = \tau^*$ when $\tau^*$ exists in the support $\left(\min\{a,b\},\sqrt{a^2+b^2}\right)$.
		\end{enumerate}
		\label{prop2}
	\end{proposition}
	
	\noindent A proof of Proposition 2 is provided in Appendix\ref{App-prop2}. Based on Proposition 2, we have an exponentially increasing function of $f_{\cos{\psi}}$ in the lower tail, but it is not always an exponentially decreasing function of $f_{\cos{\psi}}$ in the higher tail as in case 1. It can be inferred from \eqref{TaylorParametersb} and \eqref{eqworstconfsolab} that when $(x_{\rm{u}},y_{\rm{u}})\in \mathcal{C}_{\rm{w}}$ and $x_{\delta}=y_{\delta}=0$, then, $\hat{b_{\rm{\theta}}}=0$.\footnote{Refer to the discussion in Appendix\ref{App-prop2} for the definition of $\mathcal{C}_{\rm{w}}$.} To give readers a context, the set $\mathcal{C}_{\rm{w}}$ is a line (a hyperline in $\mathbb{R}^2$) at which the orientation of the UE faces the AP. In other words, when $(x_{\rm{u}},y_{\rm{u}})\in \mathcal{C}_{\rm{w}}$ the channel attenuation is small. Under this condition, the PDF of the approximation method turns into a probability mass function, i.e., it becomes a discrete RV. In this case, we observe a sudden jump in the KSD value \cite{ArdimasWCNC2018}. In fact, this anomaly only occurs when $(x_{\rm{u}},y_{\rm{u}})\in \mathcal{C}_{\rm{w}}$. However, the area of the line $\mathcal{C}_{\rm{w}}$ in $\mathbb{R}^2$ is zero and so this is a negligible part of the typical indoor room area. Furthermore, in our recent study \cite{MDSBER}, we have shown that if $(x_{\rm u},y_{\rm u})\in \mathcal{C}_{\rm w}$, the BER is less sensitive to the random orientation.

	\section{The Statistics of Channel Gain}
	In this section, we discuss the statistics of the channel gain. First, the PDF of the channel gain given the location and direction of the user is obtained in closed form. Then, the effect of random orientation change is described as a random process over time.  
	
	\subsection{PDF of Channel Gain}
	After justifying the proposed approximation for the PDF of $\cos\psi$, we can use the simple equation given in \eqref{PDFCosineTaylorTrunc} instead of the more complicated equations given in \eqref{eqfcosphigivenomegacase1} and \eqref{eqfcosphigivenomegacase} to calculate the LOS channel gain and further related derivations. Note that the PDF of the channel gain derived in this section is the conditional PDF given the location and direction of UE. Therefore, having the statistics of the user location, the joint PDF of the channel gain with respect to both UE orientation and location can be readily obtained. In the next section, we use this approach to develop an orientation-based random waypoint model to analyze mobile wireless systems.
	
	The DC gain of the LOS optical channel can be expressed as $H=\frac{H_0\cos\psi}{d^{m+2}}$ for $0\leq\psi\leq\Psi_{\rm{c}}$ and $H=0$ for $\psi\ge\Psi_{\rm{c}}$, where $H_0=\frac{(m+1)Ah^{m}}{2\pi}$. Thus, for any given position $(x_{\rm{u}},y_{\rm{u}},z_{\rm{u}})$, the PDF of $H$ can be expressed as follows:\vspace{0.0cm}
	\begin{equation}
	\label{LOSPDF1}
	f_{\rm{H}}(\hbar)=\frac{1}{h_{\rm{n}}}f_{\cos\psi}\left( \frac{\hbar}{h_{\rm{n}}}\right)+F_{\cos\psi}\left(\cos\Psi_{\rm{c}} \right)\delta(\hbar) 
	\end{equation}
	where $h_{\rm{n}}=\frac{H_0}{d^{m+2}}$ is the normalizing factor. The support range of $f_{\rm{H}}(\hbar)$ is $h_{\rm{min}}\leq\hbar\leq h_{\rm{max}}$, 
	where $h_{\rm{min}}$ and $h_{\rm{max}}$ are given as:
	\begin{equation}
	\label{hmin}
	h_{\rm{min}}=\begin{cases}
	\dfrac{H_0}{d^{m+2}}\cos\Psi_{\rm{c}},   & {\rm{if}}\ \ \  \cos\psi<\cos\Psi_{\rm c}\\
	\dfrac{H_0}{d^{m+2}}\min\{a,b\},   & {\rm{o.w}}
	\end{cases}
	\end{equation}
	\begin{equation}
	\label{hmax}
	\!\!\!\!\!\!\!\!\!\!\!\!\!\!\!\!\!\!\!\!\!\!\!\! h_{\rm{max}}=\begin{cases}
	\dfrac{H_0}{d^{m+2}}b , & {\rm{if}}\ \ \  a<0\\
	\dfrac{H_0}{d^{m+2}}\sqrt{a^2+b^2} , & {\rm{if}}\ \ \  a\geq0
	\end{cases}
	\end{equation}
	The proof of \eqref{LOSPDF1} is provided in Appendix\ref{App-D}. After some manipulations, \eqref{LOSPDF1} can be rewritten for the sitting scenario (i.e., Laplace distribution assumption for $\theta$) as:\vspace{0.0cm}		
	\begin{equation}
	\label{LOSPDF}
	\begin{aligned}
	f_{\rm{H}}(\hbar)=\frac{\exp\left(-\frac{|\hbar-\mu_{\rm{H'}}|}{b_{\rm{H'}}} \right) }{b_{\rm{H'}}\left(2-\exp\left(-\frac{h_{\rm{max}}-\mu_{\rm{H'}}}{b_{\rm{H'}}} \right)  \right) }+F_{\cos\psi}\left(\cos\Psi_{\rm{c}} \right)\delta(\hbar),
	\end{aligned}
	\end{equation}
	with the support range of $h_{\rm{min}}\leq \hbar\leq h_{\rm{max}}$; Note that in \eqref{LOSPDF}, $H'$ denotes a Laplace distribution with the following parameters:
	\begin{align}
	\label{LOSParameters}
	\mu_{\rm{H'}} &=\frac{H_0}{d^{m+2}}\left(  a \sin{\mu_{\theta}} + b \cos{\mu_{\theta}}\right) , \\
	b_{\rm{H'}} &=\frac{H_0}{d^{m+2}}b_{\rm{\theta}} |a \cos{\mu_{\theta}} - b \sin{\mu_{\theta}} | .
	\end{align}
	
	Moreover, for the special case of $b_{\rm{H}}\approx0$, we have $f_{\rm{H}}(\hbar)\cong\delta(\hbar-h_{\rm{max}})$ and for the case that $\Pr(\psi<\Psi_{\rm{c}})\approx0$, the PDF would be $f_{\rm{H}}(\hbar)\cong\delta(\hbar)$. 
	%Also for the case that $\Pr(\psi<\Psi_{\rm{c}})\cong0$ and noting that $F_{\rm{H'}}(h_{\rm{max}})\cong1$ and $h_{\rm{min}}=0$, we have  $f_{\rm{H}}(h)\cong\frac{F_{\rm{H'}}(h_{\rm{min}})}{F_{\rm{H'}}(h_{\rm{max}})}\delta(0)\cong F_{\rm{H'}}(0)\delta(0)$.   
	
	\begin{table}[t]
		\centering
		\caption{Simulation Parameters}
		\label{TableSimulationParam}
		\vspace{-8pt}
		{\raggedright
			\vspace{4pt} \noindent
			\begin{tabular}{p{130pt}|p{25pt}|p{50pt}}
				\hline
				\parbox{130pt}{\centering{\small Parameter}} & \parbox{25pt}{\centering{\small Symbol}} & \parbox{50pt}{\centering{\small Value}} \\
				\hline
				\hline
				\parbox{130pt}{\raggedright{\small LED half-intensity angle}} & \parbox{25pt}{\centering{\small $\Phi_{1/2}$}} & \parbox{50pt}{\centering{\small $60^\circ$}} \\
				\hline
				\parbox{130pt}{\raggedright{\small Receiver FOV}} & \parbox{25pt}{\centering{\small $\Psi_{\rm{c}}$}} & \parbox{50pt}{\centering{\small $90^\circ$}} \\
				\hline
				\parbox{130pt}{\raggedright{\small Physical area of a PD}} & \parbox{25pt}{\centering{\small $A$}} & \parbox{50pt}{\centering{\small $1$ cm$^2$}} \\
				\hline
				\parbox{130pt}{\raggedright{\small PD responsivity}} & \parbox{25pt}{\centering{\small $R_{\rm{PD}}$}} & \parbox{50pt}{\centering{\small $1$ A/W }} \\
				\hline
				\parbox{130pt}{\raggedright{\small Vertical distance of UE and AP}} & \parbox{25pt}{\centering{\small $h$}} & \parbox{50pt}{\centering{\small $2$ m}} \\
				\hline
				\parbox{130pt}{\raggedright{\small Transmitted optical power}} & \parbox{25pt}{\centering{\small $P_{\rm{opt}}$}} & \parbox{50pt}{\centering{\small $1$ W}} \\
				\hline
				\parbox{130pt}{\raggedright{\small Downlink bandwidth}} & \parbox{25pt}{\centering{\small $B$}} & \parbox{50pt}{\centering{\small $10$ MHz}} \\
				%\hline
				%\parbox{130pt}{\raggedright{\small Conversion factor}} & \parbox{25pt}{\centering{\small $\eta$}} & \parbox{50pt}{\centering{\small $3$ }} \\
				\hline
				\parbox{130pt}{\raggedright{\small Noise power spectral density}} & \parbox{25pt}{\centering{\small $N_0$}} & \parbox{50pt}{\centering{\small $10^{-21}$ A$^2$/Hz }} \\
				\hline
			\end{tabular}
			\vspace{2pt}
		}
		\vspace{-10pt}
	\end{table}
	
\begin{figure}
		\centering
		\begin{subfigure}[b]{0.5\columnwidth}
			\centering
			\includegraphics[width=1\columnwidth,draft=false]{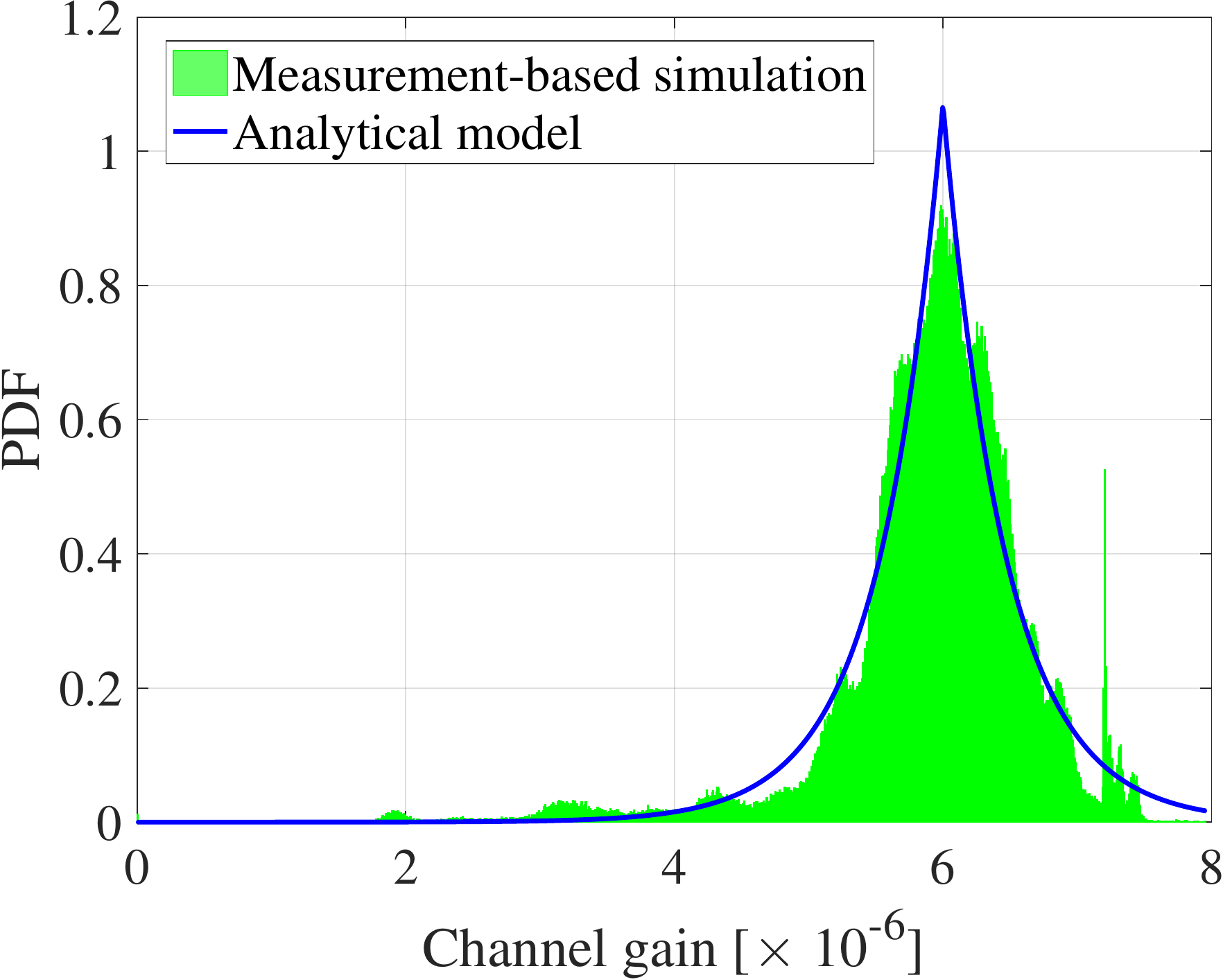}
			\caption{$x_{\rm{u}}=0$, $y_{\rm{u}}=0$}
		\end{subfigure}%
		~
		\begin{subfigure}[b]{0.5\columnwidth}
			\centering
			\includegraphics[width=1\columnwidth,draft=false]{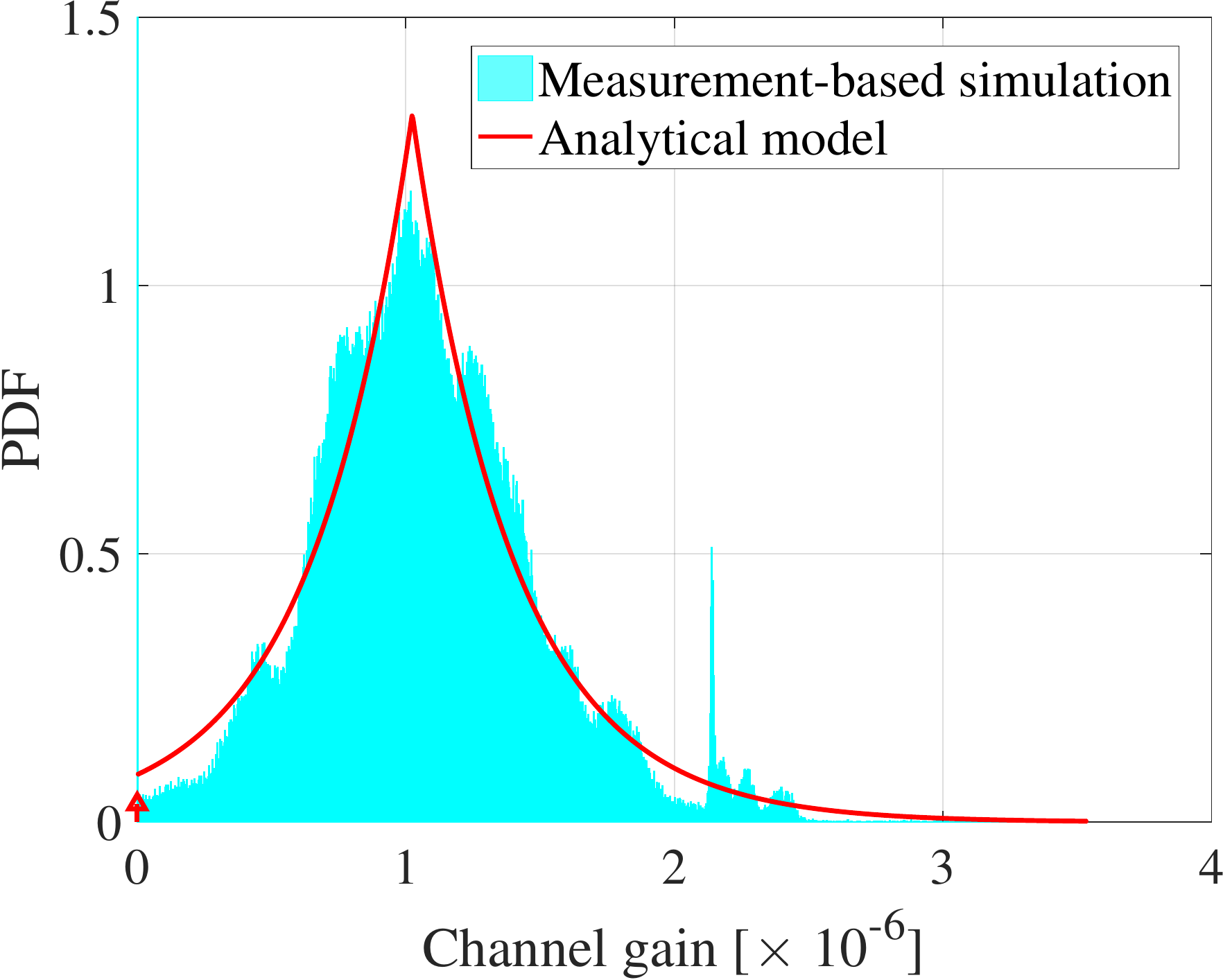}
			\caption{$x_{\rm{u}}=-1$, $y_{\rm{u}}=-1$}
		\end{subfigure}\\
		\begin{subfigure}[b]{0.5\columnwidth}
			\centering
			\includegraphics[width=1\columnwidth,draft=false]{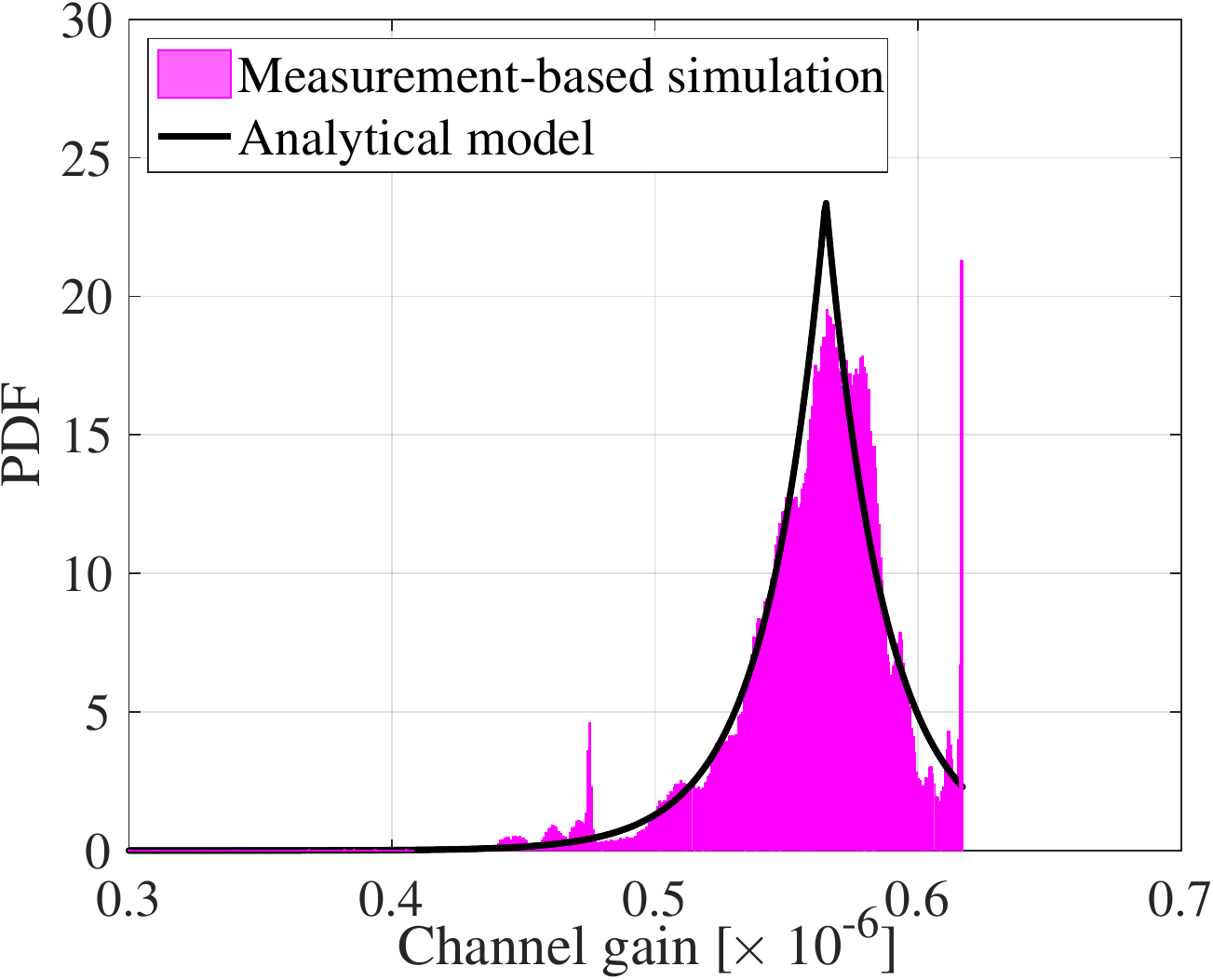}
			\caption{$x_{\rm{u}}=3$, $y_{\rm{u}}=3$}
		\end{subfigure}%
		~
		\begin{subfigure}[b]{0.5\columnwidth}
			\centering
			\includegraphics[width=1\columnwidth,draft=false]{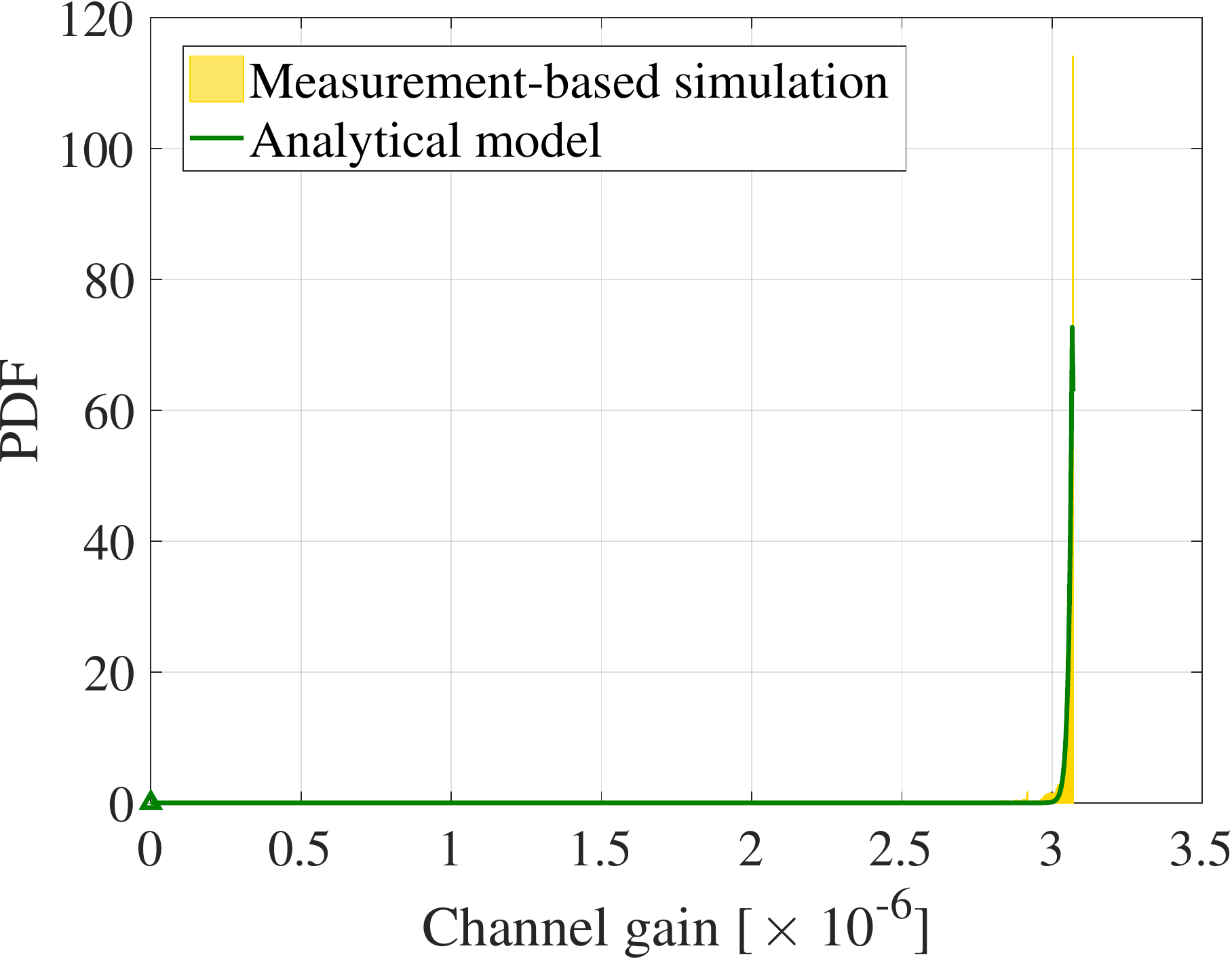}
			\caption{$x_{\rm{u}}=1.33$, $y_{\rm{u}}=1.33$}
		\end{subfigure}
		\caption{Comparison between measurement-based simulations and analytical results of PDF of channel gain for different locations in the room with $\Omega=\frac{\pi}{4}$. }
		\label{figbePDFHLOS}
		\vspace{-10pt}
	\end{figure}

	Fig.~\ref{figbePDFHLOS} represents the PDF of the LOS channel gain obtained from analytical results given in \eqref{LOSPDF} and the measurement-based simulations. The results are provided for different positions in the room with $\Omega=\frac{\pi}{4}$. The simulation parameters are provided in Table~\ref{TableSimulationParam}. The results show the accuracy of the derived analytical PDF. The magnitude of the Dirac delta term is almost zero in Fig.~\ref{figbePDFHLOS}-(a) and Fig.~\ref{figbePDFHLOS}-(c) while it is $0.0336$ and $0.006$ in Fig.~\ref{figbePDFHLOS}-(b) and Fig.~\ref{figbePDFHLOS}-(d), respectively.
	Fig.~\ref{figbePDFHLOS}-(a) and Fig.~\ref{figbePDFHLOS}-(b) illustrate two positions that correspond to $a=0$ and $a<0$, respectively while Fig.~\ref{figbePDFHLOS}-(c) and Fig.~\ref{figbePDFHLOS}-(d) present the positions associated to $a>0$. As can be seen, for $a\leq 0$, the analytical derivation of LOS channel gain and the measurement-based simulations match very well. For $a>0$, we still observe a good accommodation between analytical and simulation results and in fact the difference happens at $\hbar=h_{\rm{max}}$ as shown in Fig.~\ref{figbePDFHLOS}-(c) and Fig.~\ref{figbePDFHLOS}-(d). 
	We also observe that the distribution of the channel gain significantly changes from the Laplacian shape as $a\longrightarrow 1$ or equivalently $(x_{\rm{u}},y_{\rm{u}})$ approaches to the $\mathcal{C}_{\rm{w}}$ region as shown in Fig.~\ref{figbePDFHLOS}-(d). 
	
	\textit{SNR Statistics: }The received electrical SNR in an optical wireless channel can be obtained as:
	\begin{equation}
	\mathcal{S}=\frac{R_{\rm{PD}}^2H^2P_{\rm{opt}}^2}{N_0B},
	\end{equation}
	where $R_{\rm{PD}}$ is the PD responsivity; $P_{\rm{opt}}$ is the transmitted optical power; $N_0$ is the single sided power spectral density of noise; $B$ is the modulation bandwidth. Let's define $\mathcal{S}_0\triangleq \frac{R_{\rm{PD}}^2P_{\rm{opt}}^2}{N_0B}$, then we have $\mathcal{S}=\mathcal{S}_0H^2$. Using the fundamental theorem of determining the distribution of a random variable \cite{papoulis1985random} and noting that $H\geq0$, we have:\vspace{0.0cm}
	\begin{equation}
	\label{SNRPDF}
	\begin{aligned}		
	&f_{\mathcal{S}}(s)=\frac{f_H(\sqrt{s/\mathcal{S}_0})}{2\mathcal{S}_0\sqrt{s/\mathcal{S}_0}}
	\\=&\frac{\exp\left(-\frac{|\sqrt{s}-\sqrt{\mathcal{S}_0}\mu_{\rm{H'}}|}{\sqrt{\mathcal{S}_0}b_{\rm{H'}}} \right) }{2b_{\rm{H'}}\!\sqrt{\mathcal{S}_0s}\left(2-\exp\left(-\frac{h_{\rm{max}}-\mu_{\rm{H'}}}{b_{\rm{H'}}} \right)  \right)}\!+F_{\cos\psi}\left(\cos\Psi_{\rm{c}} \right)\delta(s),
	\end{aligned}
	\end{equation}
	with the support range of $s\in(s_{\rm{min}},s_{\rm{max}})$, where $s_{\rm{min}}=\mathcal{S}_0h_{\rm{min}}^2$ and $s_{\rm{max}}=\mathcal{S}_0h_{\rm{max}}^2$, with $h_{\rm{min}}$ and $h_{\rm{max}}$ given in \eqref{hmin} and \eqref{hmax}, respectively. %It can be shown that $c_{\delta}$ in \eqref{SNRPDF} is equal to $c_{\delta}=F_H\left(h_{\rm{min}}\right)$, where $F_H$ is the CDF of the LOS channel gain.
	
	\subsection{Random Process Model for Device Orientation}
	If a fixed orientation is assumed for the UE with an angle  $\psi_0\in\left[ 0,\Psi_{\rm{c}}\right] $, the LOS channel gain remains constant and equals to $H=\frac{H_0\cos\psi_0}{d^{m+2}}$. However, due to the random orientation of the UE, the incidence angle indeed fluctuates around the angle $\psi_0$ as observed in the experimental measurements. Thus, the LOS channel gain also varies and it can be observed as a stochastic random process (RP). Note that in a realistic scenario whether the user is in a walking or sitting position, it is fair to assume that the angle $\Omega$ (corresponding to the direction of the movement) is fairly stable and does not fluctuate with the same rate as the polar angle. We therefore first focus on the conditional probability of the channel gain given a particular direction of the movement. We then in the next section introduce a random waypoint model to consider the random effect of $\Omega$ (random change of direction) on the channel gain and thus the communication system performance.
	
\begin{figure}
		\begin{center}
			\includegraphics[width=1\columnwidth,draft=false]{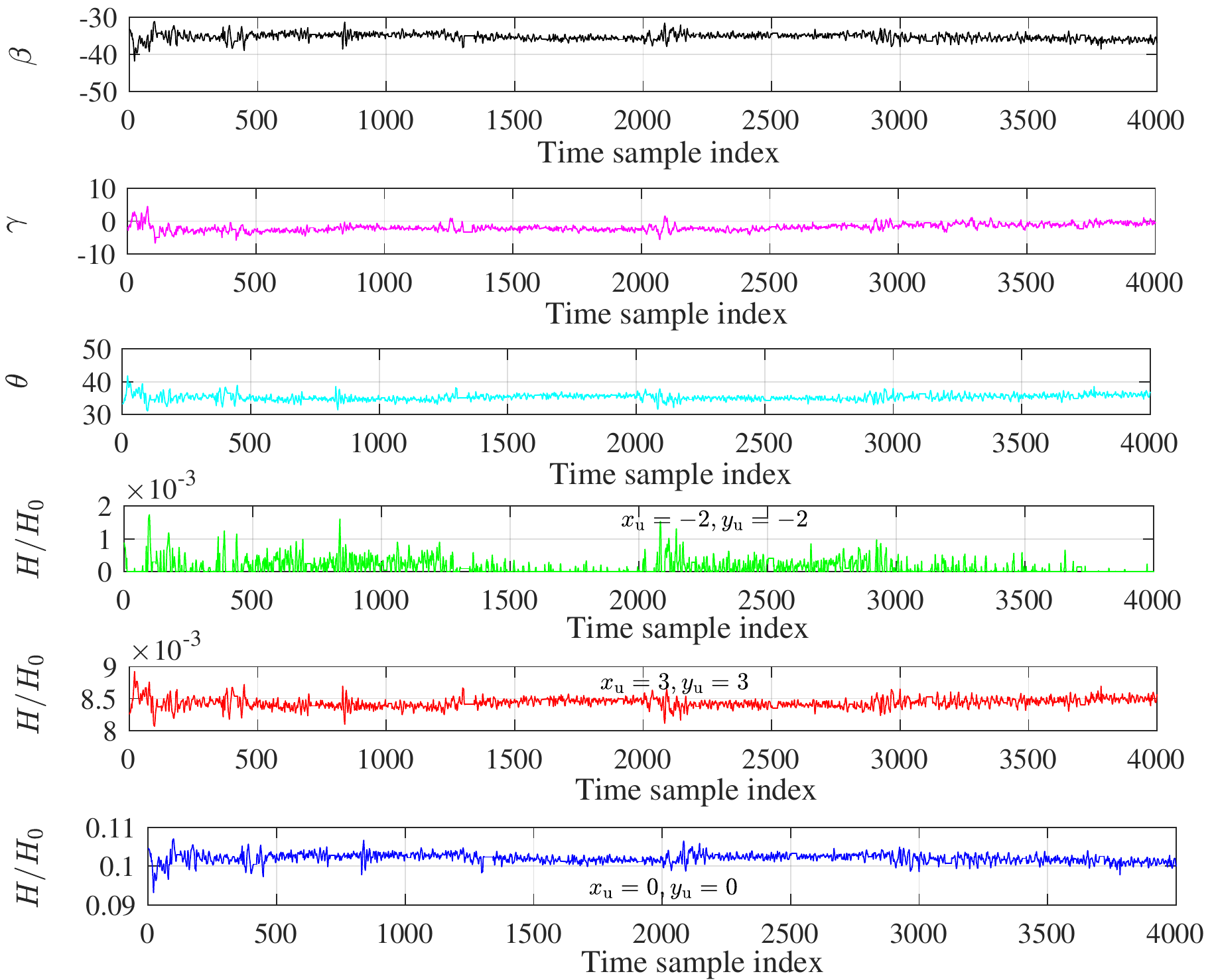}
			\caption{Representation of one ensemble of the stochastic RPs, $\beta$, $\gamma$, $\theta$ and the normalized LOS channel gain, $H/H_0$, for three UE's locations of $(-2,-2)$, $(3,3)$ and $(0,0)$. Here, $\Omega\!=\!\frac{\pi}{4}$ and $T_{\rm{s}}\!=\!13$ ms.}
			\label{figStochasticProcessOrientation}
			\vspace{-0.3cm}
		\end{center}
		\vspace{-10pt}
	\end{figure}

	Fig.~\ref{figStochasticProcessOrientation} illustrates one ensemble of the stochastic processes $\beta$, $\gamma$ and $\theta$. As can be seen, the pitch angle, $\beta$, varies around the mean value of $\mathbb{E}[\beta]=-35.81^\circ$ and the roll angle, $\gamma$, fluctuates around its mean value which is zero. This means that the user tends to hold its cellphone with the pitch angle of $-35.81^\circ$ and roll angle of zero. However, due to the random nature of users behavior, fluctuations are observed  in pitch, roll and polar angles.
	The variations in pitch and roll angles also affect the LOS channel gain. Fig.~\ref{figStochasticProcessOrientation} shows three ensembles of the normalized LOS channel gain, $\frac{H}{H_0}$, for UE's locations of $(x_{\rm{u}},y_{\rm{u}})\!=\!(-2,-2)$, $(x_{\rm{u}},y_{\rm{u}})\!=\!(3,3)$ and $(x_{\rm{u}},y_{\rm{u}})\!=\!(0,0)$ with $\Omega=\frac{\pi}{4}$.

To characterize the temporal behavior of the device random orientation as a random process, the coherence time of the channel needs to be known. Autocorrelation is one common way to calculate the coherence time of a stochastic random process. Denoting the autocorrelation of $\theta$ as $\mathcal{R}_{\rm{\theta}}(t)$, the coherence time of the polar angle stochastic process denoted by $T_{\rm{c,\theta}}$ represents the time it takes for the random process to become uncorrelated with its initial value and can be defined as $\mathcal{R}_{\rm{\theta}}(T_{\rm{c,\theta}})=0.05\mathcal{R}_{\rm{\theta}}(0)=0.05$ \cite{tse2005fundamentals}.  %$\ell_{\rm{\theta}}=\mathcal{R}_{\rm{\theta}}^{-1}(0.05)$ and $\mathcal{R}_{\rm{\theta}}^{-1}(\cdot)$ is the inverse autocorrelation function of $\theta$.   		 
As expressed in section~\ref{ExperimentalSetup}, our experimental results show that the coherence time of the angle $\theta$ is in the order of a few hundreds of milliseconds \cite{ArdimasOFDM}. However, the coherence time of the LOS channel gain highly depends on the UE's location and direction (determined by the angle $\Omega$), which is in the order of a few tens of milliseconds. Accordingly, the effect of random orientation may be modeled as a slow large-scale fading effect as the coherence time of the channel variation due to random orientation, $T_{\rm{c,H}}$, is very large compared to the typical symbol period in OWC. This implies that the channel gain can be presumed as roughly constant over a large number of transmitted symbols\cite{tse2005fundamentals}. 

	\section{Orientation-based Random Waypoint}
	\label{ORWP}
	A commonly used mobility model in simulation based studies of wireless networks is the random waypoint (RWP) model. According to the RWP mobility model, i) users choose their destinations randomly in the room area, ii) they move with a constant speed on a straight line between a source and a destination. Note that destinations are uniformly randomly distributed in the room area \cite{Bettstetter}. RWP mobility model is described as a discrete-time stochastic process. Here, we consider the RWP movement in two-dimensional space. The angle between the direction of movement and the positive direction of the $X$-axis is defined as the angle of direction. This angle is the same as $\Omega$ given in \eqref{OmegaPortrait} and \eqref{OmegaLandscape}. 
	The RWP can be mathematically denoted as an infinite sequence of triples, $ \left\{ \left( \mathbf{P}_{n-1}, \mathbf{P}_n, v_n \right) \big| \ n \in \mathbb{N} \right\} $, where $n$ stands for the $n$th movement period. The UE moves from the random waypoint ${\bf{P}}_{n-1}=(x_{n-1},y_{n-1})$ to the destination point ${\bf{P}}_{n}=(x_n,y_n)$ with speed of $v_{n}$ chosen from a random distribution of $f_v$. %Then, it pauses at the destination for a period of $\tau_{{\rm{p}},n}$. %Fig.~\ref{figRWP} illustrates the RWP mobility model and the angle of direction in a room of size $L\times L$. 
	The Euclidean distance between two consecutive waypoints, ${\bf{P}}_{n-1}$ and ${\bf{P}}_{n}$ is defined as the transition length and is given by $\mathcal{D}_n=\|{\bf{P}}_{n}-{\bf{P}}_{n-1} \|$. 
	The sequence of transition lengths $\{\mathcal{D}_1, \mathcal{D}_2, ... \}$ are independent identically distributed (i.i.d.) RVs with the PDF described in \cite{Bettstetter}, and the expected transition length of $\mathbb{E}[\mathcal{D}]=0.5214L$ in a room of size $L\times L$ m$^2$. 
	The elapsed time between two successive movements for the $n$th period can be obtained as $T_{{\rm{e}},n}=\mathcal{D}_n/v_n$. The expected value of the elapsed time is given as $\mathbb{E}[T_{\rm{e}}]=\mathbb{E}[\mathcal{D}]\mathbb{E}\left[\frac{1}{v}\right]$. 
	
	In the context of LiFi as well as mmWave cellular networks, the effect of UE's orientation on the performance of the system is significant. In fact, a significant change of UE's orientation, whether individually or combined with the mobility of the user, can lead to a handover that would not normally happen for UEs with a constant orientation. So in order to provide a framework to analyze the performance of mobile wireless networks more realistically, we need to combine the conventional RWP with the random orientation model. Therefore, the orientation-based random waypoint (ORWP) can be modeled as an infinite sequence of quadruples, $ \left\{ \left( \mathbf{P}_{n-1}, \mathbf{P}_n, v_n, \theta_n(t) \right) \big| \ n \in \mathbb{N} \right\} $, where $\theta_{n}(t)$ is a random process describing the UE's polar angle during the movement from waypoint ${\bf{P}}_{n-1}$ to waypoint ${\bf{P}}_{n}$. More discussions about how to generate these sequences are presented next and the ORWP is summarized in the Algorithm~\ref{Algorith1}.
	
	\begin{algorithm}[t]
		\caption{Orientation-based random waypoint (ORWP)}
		\label{Algorith1}
		
		\begin{algorithmic}[1]
			\STATE Initialization: $n\longleftarrow 1$; $k\longleftarrow 1$;\\ 
			denote ${\bf{P}}_{n}=(x_n,y_n)$ as the $n$th location of UE and ${\bf{P}}_{0}=(x_0,y_0)$ as the initial UE's position;\\
			$N_{\rm{r}}$ as the number of runs;\\
			$v$ as the speed of UE;\\
			$T_{\rm{c,\theta}}$ as the coherence time of the polar angle; \\
			$\mathbb{E}[\theta]$ and $\sigma_{\theta}^2$ as the mean and variance of Gaussian RP;
            
			\FOR{  $k=1:N_{\rm{r}}$ }
			%\vspace{1mm}
			\STATE Choose a random position ${\mathcal{P}}_{k}=(x_k,y_k)$
			%\vspace{1mm}
			\STATE Compute $\mathcal{D}_k=\|{\mathcal{P}}_{k}-\mathcal{P}_{k-1} \|$
			%\vspace{1mm}
			\STATE Compute $\Omega=\tan^{-1}\left(\frac{y_k-y_{k-1}}{x_k-x_{k-1}} \right) $
			\STATE $t_{\rm{move}}\longleftarrow 0$; 
			%\STATE $t_{\rm{0}}\longleftarrow 0$; 
			%\vspace{1mm}
			\WHILE{$t_{\rm{move}}\leq\frac{\mathcal{D}_k}{v}$}
			%\vspace{1mm}
			%\vspace{1mm}
			\STATE Compute ${\bf{P}}_{n}=(x_n,y_n)$ with \\ $x_n=x_{n-1}+vT_{\rm{c,\theta}} \cos\Omega$ and $y_n=y_{n-1}+vT_{\rm{c,\theta}} \sin\Omega$
			%\vspace{1mm}
			\STATE Generate $\theta_n$ based on the AR(1) model
			%\vspace{1mm}
			\STATE Return $({\bf{P}}_{n-1},{\bf{P}}_{n},v,\theta_n)$ as ORWP specifications
			\STATE $n\longleftarrow n+1$
			\STATE $t_{\rm{move}}\longleftarrow t_{\rm{move}}+T_{\rm{c,\theta}}$
			%\vspace{1mm}
			\ENDWHILE
			%\vspace{1mm}
			\IF{$t_{\rm{move}}\neq \frac{\mathcal{D}_k}{v}-T_{\rm{c,\theta}}$}
			%\vspace{1mm}
			\STATE Generate $\theta_n$ based on the AR(1) model
			%\vspace{1mm}
			\STATE ${\bf{P}}_{n}\longleftarrow {\mathcal{P}}_{k}$
			%\vspace{1mm}
			\STATE Return $({\bf{P}}_{n-1},{\bf{P}}_{n},v,\theta_n)$ as ORWP specifications
			\STATE $n\longleftarrow n+1$
			\ENDIF
			%\STATE Choose a random pause time $T_{{\rm{p}},k}$ from a random distribution of $f_{T_{{\rm{p}}}}$
			%\ENDIF
			%\STATE $t_{\rm{pause}}\longleftarrow 0$
			%\vspace{1mm}
			%\WHILE{$t_{\rm{pause}}\leq T_{{\rm{p}},k}$}
			%\vspace{1mm}
			%\STATE Generate $\theta_n$ based on the AR(1) model
			%\vspace{1mm}
			%\STATE ${\bf{P}}_{n}\longleftarrow {\bf{P}}_{n-1}$
			%\vspace{1mm}
			%\STATE Return $({\bf{P}}_{n-1},{\bf{P}}_{n},v,T_{{\rm{p}},k},\theta_n)$ as ORWP specifications
			%\STATE $n\longleftarrow n+1$
			%\STATE $t_{\rm{pause}}\longleftarrow t_{\rm{pause}}+T_{\rm{c,\theta}}$
			%\vspace{1mm}
			%\ENDWHILE
			%\IF{$t_{\rm{pause}}\neq T_{{\rm{p}},k}-T_{\rm{c,\theta}}$}
			%\vspace{1mm}
			%\STATE Generate $\theta_n$ based on the AR(1) model
			%\vspace{1mm}
			%\STATE ${\bf{P}}_{n}\longleftarrow {\bf{P}}_{n-1}$
			%\vspace{1mm}
			%\STATE Return  $({\bf{P}}_{n-1},\!{\bf{P}}_{n},\!v,\!T_{{\rm{p}},k},\theta_n)$ as ORWP specifications
			%\STATE $n\longleftarrow n+1$
			%\ENDIF
			\STATE $k\longleftarrow k+1$
			%\vspace{1mm}
			\ENDFOR
			%\vspace{1mm}
		\end{algorithmic}
	\end{algorithm}
	
	\subsection{Correlated Gaussian Random Process}
	As shown in section~\ref{ExperimentalSetup}, the polar angle for walking activities follows a Gaussian distribution. The experimental measurements also illustrate that the adjacent samples of the RP, $\theta$, are correlated. Therefore, in order to incorporate the orientation with the RWP mobility model, it is required to generate a correlated Gaussian RP that statistically matches the experimental measurements. 
	Possible ways of generating a correlated Gaussian RP can be found in %\cite{bartosch2001generation,kay1981efficient,fox1988fast,CorrelatedGaussTrans,CorrelatedGaussProceedings} 
\cite{fox1988fast,CorrelatedGaussTrans} and references therein.  A simple method to generate a correlated Gaussian RP is to pass a white noise process through a linear time-invariant (LTI) filter, e.g., using a linear autoregressive (AR) model. Let $w[n]$ denote the white noise process, then, after passing it through the LTI filter, the $n$th time sample of the correlated Gaussian RP, $\theta[n]$, is given as:
% * <majid.safari@ed.ac.uk> 2018-05-16T07:17:39.730Z:
% 
% > references therein
% Why so many references? reduce them to max two that are most useful.
% 
% ^.
	\begin{equation}
	\theta[n]=c_0+\sum_{i=1}^{p}c_i\theta[n-i]+w[n],
	\end{equation}
	where $c_0$ determines the biased level and $c_i$ for $i=1,\dots,p$ are constant factors of the AR order of $p$, AR($p$). Note that AR($p$) contains $p+2$ unknown parameters including: $c_0,c_1,\dots,c_p,\sigma_w^2$, where $\sigma_w^2$ is the variance of white noise RP, $w$. In this study, we focus on matching the generated random process to the moments and the coherence time of the polar angle measured experimentally rather than its exact ACF. Therefore, it is sufficient to assume $p=1$ and consider first-order AR model to generate the correlated Gaussian RP. Thus, the $n$th sample of the AR(1) model can be expressed as:
	\begin{equation}
	\label{AR1}
	\theta[n]=c_0+c_1\theta[n-1]+w[n],
	\end{equation}
	where $c_1$ should meet the condition $|c_1|<1$ to guarantee the RP, $\theta$, is wide-sense stationary. Noting that for AR(1), the mean, variance and ACF are given as \cite{box2015time}:
	\begin{equation*}
	\label{MeanAR}
	\mathbb{E}[\theta]=\frac{c_0}{1-c_1}, \ \ \ \ \ \ \ \sigma_{\theta}^2=\frac{\sigma_w^2}{1-c_1^2}, \ \ \ \ \ \ \ \mathcal{R}_{\theta}(\ell)=c_1^\ell.
	\end{equation*}
	Using the above equations and noting that $\mathcal{R}_{\rm{\theta}}(\ell_{\rm{\theta}}=\frac{T_{\rm{c,\theta}}}{T_{\rm{s}}})=0.05$ where $T_s$ is the sample time, we have:
	\begin{equation}
	c_0=(1-c_1)\mathbb{E}[\theta],\ \ \ \ \ \ \ \sigma_w^2=(1-c_1^2)\sigma_{\theta}^2, \ \ \ \ \ \ \ 
	c_1=0.05^{\frac{T_{\rm{s}}}{T_{\rm{c,\theta}}}}. 
	\end{equation}
	Once the parameters of the AR(1) model are determined, the $n$th time sample of the correlated Gaussian RP, $\theta$, can be specified according to \eqref{AR1}. 
	Using the method described above, the ORWP is presented in the Algorithm~\ref{Algorith1}.

	\subsection{Use Case: Handover Rate}
	Here, we investigate the effect of the UE's orientation on the handover rate as a case study. 
	In fact, one of the key metrics in cellular network design is the handover rate which is defined as \cite{MDSHandover}: \vspace{0.0cm}
	\begin{equation}
	\mathcal{H}=\frac{\mathbb{E}[N_{{\rm{h}}}]}{\mathbb{E}[T_{{\rm{e}}}]}=\frac{\mathbb{E}[N_{{\rm{h}}}]}{\mathbb{E}[\mathcal{D}]\mathbb{E}\left[\frac{1}{v}\right]},
	\end{equation}
	where $\mathbb{E}[N_{{\rm{h}}}]$ is the expected number of handovers during the expected elapsed time, $\mathbb{E}[T_{{\rm{e}}}]$. Under the assumption of fixed $v_n=v$, we have $\mathbb{E}[T_{\rm{e}}]=\frac{1}{v}\mathbb{E}[\mathcal{D}]$. For a room of length $L$ the expected number of handover can be obtained as \cite{MDSHandover}:\vspace{-0.1cm}
	\begin{equation}
	\mathbb{E}[N_{{\rm{h}}}]=\frac{1}{L^4}\int\limits_{-\frac{L}{2}}^{\frac{L}{2}}\int\limits_{-\frac{L}{2}}^{\frac{L}{2}}\int\limits_{-\frac{L}{2}}^{\frac{L}{2}}\int\limits_{-\frac{L}{2}}^{\frac{L}{2}}N_{{\rm{h}}}((x,y)|(x_0,y_0))\ {\rm{d}}x{\rm{d}}y{\rm{d}}x_0{\rm{d}}y_0, 
	\end{equation}
	where the UE is assumed to move from the initial RWP $(x_0,y_0)$ to the destination point $(x,y)$. 
	
	\begin{figure}
		\begin{center}
			\resizebox{0.9\linewidth}{!}{\includegraphics{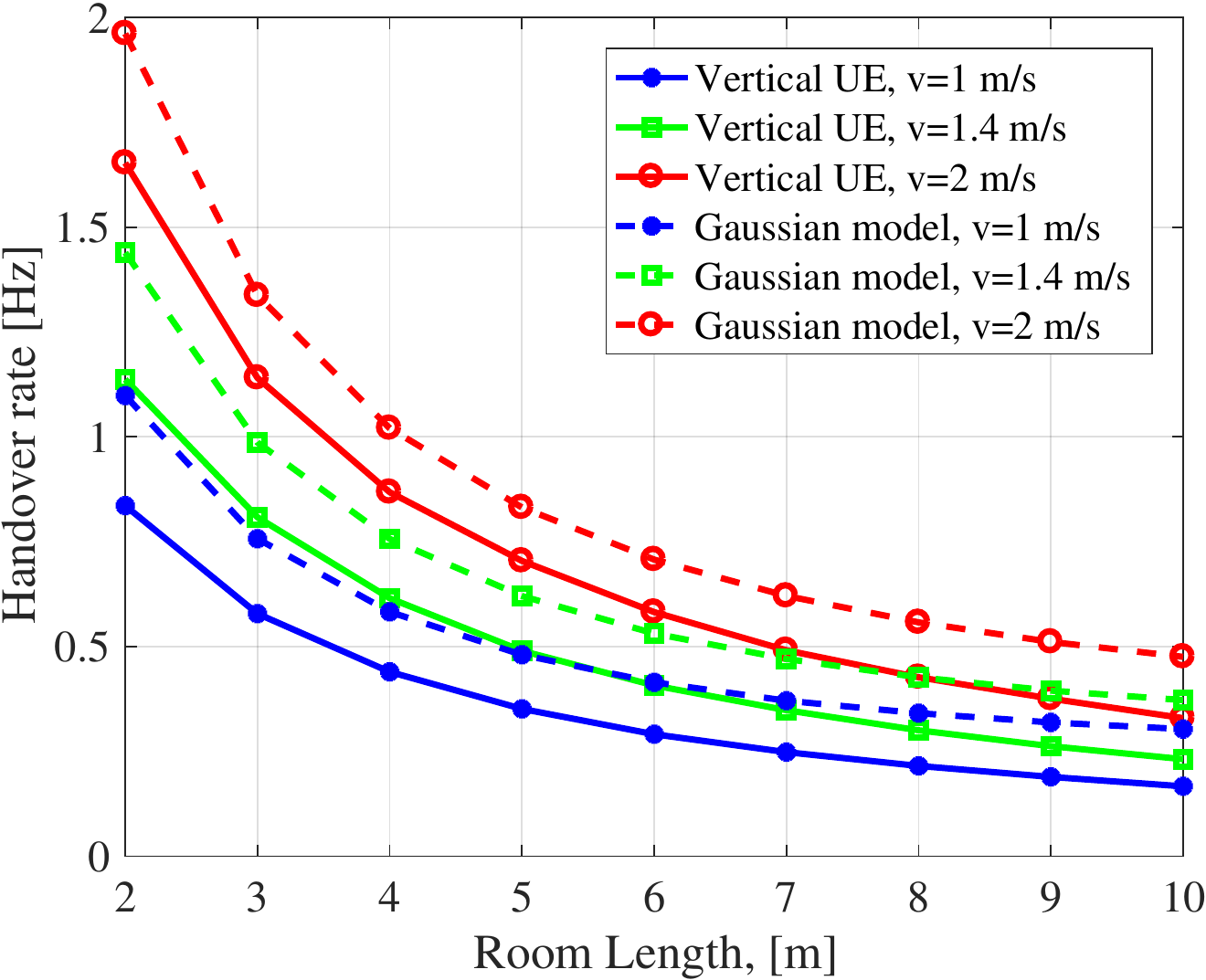}}
			\caption{Comparison of handover rate as a function of room length, $L$, for vertically upward UE and a UE with correlated Gaussian RP for $\theta$.}
			\label{fighandoevrRate}
		\end{center}
		\vspace{-15pt}
	\end{figure}
	%Now we present the results of handover rate for orientation-based RWP in a room with four APs. With four APs in the setup, the network area is divided into four quadrants with one AP at the center of each quadrant. In each quadrant, the UE is assumed to be initially connected to the corresponding AP denoted as AP$_j$ for $j = 1, 2, 3, 4$.
	Fig.~\ref{fighandoevrRate} shows the Monte-Carlo simulations of handover rate as a function of room length for a UE moving at a speed of $v=1$ m/s, $v=1.4$ m/s and $v=2$ m/s. These three speed values are chosen around the average human walking speed \cite{bohannon1997comfortable}. In this simulation setup, four APs are assumed that they divide the network area into four separate quadrants (attocells) with one AP at the center of each attocell. The UE is assumed to be initially connected to the corresponding AP denoted as AP$_j$ for $j \in \{1, 2, 3, 4\}$.
	We compared a vertically upward UE (shown by the solid lines in Fig.~\ref{fighandoevrRate}) with a UE that follows the orientation-based RWP mobility model in which the random orientation of the UE is generated based on the correlated Gaussian RP (shown by the dotted line). The mean and variance of the generated samples are chosen from Table~\ref{tabtheta} for walking activities, which are $\mathbb{E}[\theta]=29.67$ and $\sigma_{\theta}^2=7.78$. Furthermore, the coherence time of $T_{\rm{c,\theta}}=130$ ms as reported in \cite{ArdimasOFDM} for walking activities is chosen.  
% * <majid.safari@ed.ac.uk> 2018-05-16T07:33:22.293Z:
% 
% > (shown by the dotted line)
% You need to mention the coherence time used as well as the moments used to generated RP.  
% 
% Also what is the unit of hand over rate in the figure?
% 
% ^.
	As observed from Fig.~\ref{fighandoevrRate}, the handover rate is decreasing overall with an increase in the network dimensions. On the other hand, as the UE's speed increases, the handover rate increases. This is more remarkable when the network dimensions are smaller. More importantly, we observe that the effect of random orientation significantly increases the value of handover rate in all conditions while smaller network sizes are more influenced by this effect. That is, as the room length is smaller, the gap between the random orientation and vertically upward scenarios is greater.  	
	
	\section{Conclusions}
	A new model for device orientation based on the experimental measurements is proposed. The experimental measurements are taken from forty participants creating $222$ datasets for orientation. It is shown that the PDF of the polar angle follows a Laplace distribution for static users while it is better fitted to a Gaussian distribution for mobile users. The exact PDF of the cosine of the incidence angle is derived with analytical discussions. We also proposed an approximation PDF using the truncated Laplace distribution and the accuracy of this approximated method was confirmed by using the KSD test. The LOS channel gain statistics are calculated and it is described as a random process for which the coherence time was discussed. The influence of the random orientation on received SNR of OWCs is evaluated. By means of AR(1) model, an orientation-based random waypoint mobility model is proposed by considering the random orientation of the UE during the user's movement. The performance of the mobility model is assessed on the handover rate and it is shown that it is important to take the random orientation into account. 
	
	\section*{Acknowledgement}
	The first and second authors contributed equally to this study. %Authors gratefully acknowledge support by the UK EPSRC under grant EP/L020009/1 (TOUCAN Project). Harald Haas also acknowledges the financial support of his research by the Wolfson Foundation and the Royal Society. 
	
	\appendices
	\section*{Appendix}
	\subsection{Derivations of \eqref{eqfcosphigivenomegacase1} and \eqref{eqfcosphigivenomegacase}}
	\label{App-A}
	
	To derive the exact PDFs as expressed in \eqref{eqfcosphigivenomegacase1} and \eqref{eqfcosphigivenomegacase}, we rely on the following identity:
	\begin{align}\label{eqcosphiidentity}
	\cos{\psi} \!=\! a \sin{\theta} + b \cos{\theta} =\! \sqrt{a^2 + b^2} \ssin{\theta + \text{atan}2\left(b,a\right)},
	\end{align}
	\noindent Since $0 < \theta < \frac{\pi}{2}$ and $b$ is assumed to always be positive, i.e., the APs are always located above the UEs, the monocity of \eqref{eqcosphiidentity} depends on the sign of $a$. If $a$ is negative, \eqref{eqcosphiidentity} is always a monotonically decreasing function of $\theta$. Meanwhile, if $a$ is not negative, \eqref{eqcosphiidentity} is always a concave downward function of $\theta$. Therefore, we classify the derivations into two cases, namely \textbf{case 1} for $a < 0$ and $b > 0$ and \textbf{case 2} for $a \geq 0$ and $b > 0$.
	
	\subsubsection{For $a < 0$ (\textbf{Case 1})} 
	For case 1, to get the transformed distribution of $\cos{\psi}$, we need the inverse function of \eqref{eqcosphiidentity} as denoted by:
	\begin{align}\label{eqinvcase1}
	\theta = -\sin^{-1}\left( \frac{\cos{\psi}}{\sqrt{a^2+b^2}} \right)-\tan^{-1}\left(\frac{b}{a} \right), 
	\end{align}
	\noindent for $a < \cos{\psi} < b$. The Jacobian of the transformation under \eqref{eqcosphiidentity} is $|J| = 1/\sqrt{a^2+b^2-\cos^{2} \phi }$. Combining the Jacobian of transformation, \eqref{eqpdftheta}, and \eqref{eqinvcase1}, we get \eqref{eqfcosphigivenomegacase1}.
	
	\subsubsection{For $a \geq 0$ (\textbf{Case 2})} 
	For case 2, we need the intervals of the domain of $\cos{\psi}$ such that the transformation \eqref{eqcosphiidentity} is one-to-one. Using \eqref{eqcosphigivenomega} and \eqref{eqthetastar}, we have two intervals which are $g(0) < \cos{\psi} < g(\theta^*)$ and $g(\frac{\pi}{2}) < \cos{\psi} \leq g(\theta^*)$. The inverse functions are denoted by:\vspace{0.0cm}
	\begin{align}\label{eqinvcase2}
	\theta = 
	\begin{cases}
	\sin^{-1}\left( \frac{\cos{\psi}}{\sqrt{a^2+b^2}} \right)-\tan^{-1}\left(\frac{b}{a} \right),& g(0) < \cos{\psi} < g(\theta^*)\\ 
	\pi-\sin^{-1}\left( \frac{\cos{\psi}}{\sqrt{a^2+b^2}} \right)-\tan^{-1}\left(\frac{b}{a} \right),& g(\frac{\pi}{2}) < \cos{\psi} \leq g(\theta^*).
	\end{cases}
	\end{align}
	\noindent Note that both the inverse functions have the same the Jacobian of the transformation which is $|J| = 1/\sqrt{a^2+b^2-\cos^{2} \phi }$. Combining the Jacobian of transformation, \eqref{eqpdftheta}, and \eqref{eqinvcase2}, we get \eqref{eqfcosphigivenomegacase}.
	
	\subsection{Proof of Proposition~1}
	\label{App-prop1}
	\begin{proof}
		To prove all characteristics in Proposition 1, let's first define some critical points that make $f_{\cos{\psi}}$ undefined. The possible points are $\pm \sqrt{a^2+b^2}$. For $a < 0$ and $b > 0$, it is straightforward to see the following inequality:
		\begin{align*}
		-\sqrt{a^2+b^2} < a < 0 < b < \sqrt{a^2+b^2}.
		\end{align*}
		Therefore, these critical points are outside the support of $f_{\cos{\psi}}$. Considering the absolute value term in \eqref{eqfcosphigivenomegacase1}, the other critical point called $\tau^*$ is given by:
		\begin{align} \label{eqtaustarcase1}
		\tau^* = \sqrt{a^2+b^2} \ssin{-\tan^{-1}\left(\frac{b}{a} \right)-\mu_{\theta}}.
		\end{align}
		This point leads to a peak for $f_{\cos{\psi}}$ similar to the peak of the Laplace or Gaussian distributions. It is straightforward to see that $a < \tau^* < b$, if $0 < \mu_{\theta} < \frac{\pi}{2}$. Using \eqref{eqtaustarcase1}, we have: 
		\begin{align*}
		\lim_{\tau \to \tau^{*+}} f_{\cos{\psi}}(\tau) \!=\! \lim_{\tau \to \tau^{*-}} f_{\cos{\psi}}(\tau) \!=\! f_{\cos{\psi}}(\tau^*)
		\!=\! \frac{1}{2 b_{\rm{\theta}} \sqrt{a^2 + b^2 - \tau^*}}.
		\end{align*}
		
		\noindent Hence, this proves that $f_\cPsi$ is well defined for $a < \tau < b$ and continuous at $\tau = \tau^*$. 
		
		To prove the first and the third characteristics, let's denote the following:
		\begin{align}
		\frac{d}{d \tau} f_{\cos{\psi}}(\tau) = 
		\begin{cases}
		C_1(\tau) \widetilde{f_1} (\tau), &~a < \tau < \tau^* \\
		C_2(\tau) \widetilde{f_2} (\tau), &~\tau^* < \tau < b,
		\end{cases}
		\end{align}
		\noindent where
		\begin{align*}
		\widetilde{f_1} (\tau) &= \exp\left( \frac{\sin^{-1}\left( \frac{\tau}{\sqrt{a^2 + b^2}} \right) + \tan^{-1}\left( \frac{b}{a} \right) + \mu_{\theta}}{b_{\rm{\theta}}} \right),\\
		\widetilde{f_2} (\tau) &= \exp\left( -\frac{\sin^{-1}\left( \frac{\tau}{\sqrt{a^2 + b^2}} \right) + \tan^{-1}\left( \frac{b}{a} \right) + \mu_{\theta}}{b_{\rm{\theta}}} \right),
		\end{align*}
		\noindent and
		\begin{align}\label{eqcoeffrateofchange}
		C_1(\tau) &= \frac{a^2+b^2 + \tau  \left(-\tau+b_{\rm{\theta}} \sqrt{a^2+b^2-\tau ^2} \right)}{2 b_{\rm{\theta}}^2 \left(a^2+b^2-\tau ^2\right)^2} \nonumber \\
		C_2(\tau) &= -\frac{a^2+b^2 - \tau  \left(\tau+b_{\rm{\theta}} \sqrt{a^2+b^2-\tau ^2} \right)}{2 b_{\rm{\theta}}^2 \left(a^2+b^2-\tau ^2\right)^2}.
		\end{align}
\noindent Note that we base this on the Laplace distribution to derive $\widetilde{f_1}$ and $\widetilde{f_2}$ for simplicity. For the Gaussian case, they have similar forms, but the argument inside the exponential function is squared and both $\widetilde{f_1}$ and $\widetilde{f_2}$ are multiplied by a function of: $$ \frac{-\sin^{-1}\left( \frac{\tau}{\sqrt{a^2 + b^2}} \right) - \tan^{-1}\left( \frac{b}{a} \right) + \mu_{\theta}}{\sigma^2_{\rm{G}}}.$$ Note that this does not change the fact that the function $f_{\cos{\psi}}$ is still a monotonically increasing or decreasing function in an exponential manner.		
		
        Since $\widetilde{f_1} (\tau)$ and $\widetilde{f_2} (\tau)$ are both positive and monotonic functions in their domains of interest, we need to focus on the coefficients $C_1(\tau)$ and $C_2(\tau)$ as they determine the monotonicity of $f_{\cos{\psi}}$.
		
		Let's focus on the scenario where $x_{\text{a}} \neq x_{\text{u}}$ or $y_{\text{a}} \neq y_{\text{u}}$. Since the denominators are always positive, we only need to observe the numerators, so it is straightforward to have:
		\begin{align}\label{eqineqC1C2case1}
		b_{\rm{\theta}} &< -\frac{b}{a} \implies C_1(\tau) > 0 \nonumber \\
		b_{\rm{\theta}} &< -\frac{a}{b} \implies C_2(\tau) < 0. 
		\end{align}
		\noindent Satisfying the above inequalities guarantees that $f_{\cos{\psi}}(\tau)$ is a monotonically increasing function for $a < \tau < \tau^*$ and a monotonically decreasing function for $\tau^* < \tau < b$. The consequence of satisfying \eqref{eqineqC1C2case1} is that $\tau^*$ is a global maximum of $f_{\cos{\psi}}$ in the interval $(a,b)$.
		This completes the proof of Proposition 1.
	\end{proof}
	
	\subsection{Proof of Proposition~2}
	\label{App-prop2}
	\begin{proof}
		As in Appendix-\ref{App-prop1}, we have following inequality:
		\begin{align}
		-\sqrt{a^2+b^2} < \min\{a,b\}.
		\end{align}
		\noindent It means that for the lower tail, the support does not include the term $-\sqrt{a^2+b^2}$, which makes $f_\cPsi$ undefined. To see the higher tail, it is not straightforward and we need to introduce a region $\mathcal{C}_{\rm{w}}$ so that for given $\left(x_{\text{a}},y_{\text{a}},z_{\text{a}} \right)$, $z_{\text{u}}$, and $\Omega$, the CDF of $\cos{\psi}$ is almost $1$ in the vicinity of maximum value of $\cos{\psi}$. Thus, $\mathcal{C}_{\rm{w}}$ can be formalized as:  
		\begingroup\makeatletter\def\f@size{9.1}\check@mathfonts
		\def\maketag@@@#1{\hbox{\m@th\large\normalfont#1}}% 
		\begin{align*}
		\mathcal{C}_{\rm{w}} &\triangleq \argmax_{ (x_{\text{u}},y_{\text{u}}) } \PX{\max_{\theta}(\cos{\psi}) - \epsilon < \cos{\psi} \leq  \max_{\theta}(\cos{\psi})}\\
		&= \left\{ (x_{\text{u}},y_{\text{u}}) ~\Big|~ x_{\text{u}}=x'_{\text{u}}, y_{\text{u}}=y'_{\text{u}} ~\text{and}~ \tan\Omega = -\frac{x_{\delta}}{y_{\delta}} \right\},
		\end{align*}\endgroup \vspace{-0.2cm}
		%\addtocounter{equation}{}
		\begin{align}\label{eqworstconfsol}
		x'_{\text{u}} &= x_{\text{a}} + \left(\frac{z_{\text{a}}-z_{\text{u}}}{\tan\left( \frac{\pi}{2}-\mu_{\theta} \right)} \right) \cos{\Omega} + x_{\delta} \nonumber \\
		y'_{\text{u}} &= y_{\text{a}} + \left(\frac{z_{\text{a}}-z_{\text{u}}}{\tan\left( \frac{\pi}{2}-\mu_{\theta} \right)} \right) \sin{\Omega} + y_{\delta}.
		\end{align} 
		\noindent for small $\epsilon > 0$. Here, $x_{\delta}$ and $y_{\delta}$ are two auxiliary variables to move the UE position along the line $\mathcal{C}_{\rm{w}}$ in the room.  Note that $\mathcal{C}_{\rm{w}}$ for $x_\delta = y_\delta = 0$ can be expressed in terms of the coefficients $a$ and $b$ as:
		\begin{align}\label{eqworstconfsolab}
		\mathcal{C}_{\rm{w}} = \left\{ (x_{\text{u}},y_{\text{u}}) ~\Big|~ a = \sin{\mu_{\theta}}, b = \cos{\mu_{\theta}} \right\}.
		\end{align}
		
		Considering the absolute value term in \eqref{eqfcosphigivenomegacase}, the critical point $\tau^*$ is given as: $$\tau^* = \sqrt{a^2+b^2} \ssin{\tan^{-1}\left(\frac{b}{a} \right) + \mu_{\theta}}.$$ It is obvious to see that if $0 < \mu_{\theta} < \frac{\pi}{2}$, $\min\{a,b\} < \tau^* \leq \sqrt{a^2+b^2}$, and it is equal when $(x_{\rm{u}},y_{\rm{u}})\in \mathcal{C}_{\rm{w}}$. Therefore, unlike the case 1 ($a < 0$), $\tau^*$ may not exist in the interval $\left(\min\{a,b\}, \sqrt{a^2+b^2} \right)$. It is when $\tau^* = \sqrt{a^2+b^2}$ which is outside of the support of $f_{\cos{\psi}}$, and the case $\tau^* = \sqrt{a^2+b^2}$ only happens when $(x_{\rm{u}},y_{\rm{u}})\in \mathcal{C}_{\rm{w}}$. Hence, this proves the first characteristic in Proposition 2. 
		
		If $(x_{\rm{u}},y_{\rm{u}})\notin \mathcal{C}_{\rm{w}}$, we have $\min\{a,b\} < \tau^* < \sqrt{a^2+b^2}$ and a continuity at $\tau^*$, i.e.,:
		\begin{align*}
		\lim_{\tau \to \tau^{*+}} f_{\cos{\psi}}(\tau) &= \lim_{\tau \to \tau^{*-}} f_{\cos{\psi}}(\tau) = f_{\cos{\psi}}(\tau^*)
		= \frac{2}{2 b_{\rm{\theta}} \sqrt{a^2 + b^2 - \tau^*}}.
		\end{align*}
		\noindent This proves the fifth characteristic in Proposition 2. In addition, we have the following identity:
		\begin{align*}
		&\lim_{\tau \to \sqrt{a^2+b^2}} f_{\cos{\psi}}(\tau) = \infty,
		\end{align*}
		\noindent which says that for case 2, the end tail of $f_{\cos{\psi}}$ always tends to infinity. Therefore, discussion about the rate of change in case 2 is as not straightforward as that in case 1. 
		
		The function $f_{\cos{\psi}}$ in \eqref{eqfcosphigivenomegacase} can be rewritten as a combination of the following two expressions:
		\begin{align*}
		\widetilde{f}_{a}(\tau) = \frac{f_{\theta}\left( \sin^{-1}\left( \frac{\tau}{\sqrt{a^2 + b^2}} \right) -\tan^{-1}\left( \frac{b}{a} \right)  \right)}{\sqrt{a^2+b^2-\tau^2}},
		\end{align*}\vspace{-0.2cm}
		\begin{align*}
		\widetilde{f}_{b}(\tau) = \frac{f_{\theta}\left( -\sin^{-1}\left( \frac{\tau}{\sqrt{a^2 + b^2}} \right) -\tan^{-1}\left( \frac{b}{a} \right) + \pi \right)}{\sqrt{a^2+b^2-\tau^2}}.
		\end{align*}
		\noindent Since the derivatives of $\widetilde{f}_{a}(\tau)$ and $\widetilde{f}_{b}(\tau)$ give the same coefficients as those in the Appendix-\ref{App-prop1}, we are interested only in the coefficients as expressed in \eqref{eqcoeffrateofchange}. However, in case 2, $C_1(\tau)$ is used for the support $\min\{a,b\} < \tau < \tau^*$ and $C_2(\tau)$ is used for the support $\tau^* < \tau < \sqrt{a^2+b^2}$.
		
		For $\min\{a,b\} < \tau < \tau^*$, $C_1(\tau)$ is always positive. It means that we always have a monotonic increasing function of $f_{\cos{\psi}}$ with the support of $\min\{a,b\} < \tau < \tau^*$. Note that for the case $(x_{\rm{u}},y_{\rm{u}})\in \mathcal{C}_{\rm{w}}$, we have $\tau^* = \sqrt{a^2+b^2}$ by definition of $\mathcal{C}_{\rm{w}}$ in \eqref{eqworstconfsolab}. In this case, we have only a monotonic increasing function that goes to infinity at $\sqrt{a^2+b^2}$.
		
		For $\tau^* < \tau < \sqrt{a^2+b^2}$, let's first define 
		\begin{align*}
		\tau_d = \sqrt{\frac{a^2+b^2}{1+b_{\rm{\theta}}^2}},
		\end{align*}
		\noindent which is the point where the coefficient $C_2(\tau)$ starts changing from negative to positive values. This expression is obtained as in \eqref{eqineqC1C2case1} by only observing the sign of the numerator of $C_2(\tau)$. Then, we have the following result for $C_2(\tau)$:
		\begin{align*}
		\tau^* < \tau_d < \sqrt{a^2+b^2} \implies & C_2(\tau) < 0 ~\text{for}~ \tau^* < \tau < \tau_d \text{,} ~\text{and}\\
		&C_2(\tau) > 0 ~\text{for}~ \tau_d < \tau < \sqrt{a^2+b^2}.
		\end{align*}
		\noindent If $\tau_d < \tau^*$, $C_2(\tau)$ is always positive meaning that $f_{\cos{\psi}}$ is a monotonically increasing function. This proves the second to the fourth characteristics in Proposition 2 and completes the proof of Proposition 2. 
	\end{proof}

	\subsection{Proof of \eqref{LOSPDF1}}
	\label{App-D}
	\begin{proof}
		The CDF of the LOS channel gain, $H=H_0\cos\psi/d^{m+2}\ {\rm{rect}}\left( \frac{\psi}{\Psi_{\rm c}}\right)$ is given as:
		\begin{equation}
		\begin{aligned}
		\label{CDFH}
		F_H(\hbar)&=\Pr\left\lbrace H_0\cos\psi/d^{m+2}\ {\rm{rect}\left( \frac{\psi}{\Psi_{\rm c}}\right) } \leq\hbar \right\rbrace\\
		&=\Pr\{H_0\cos\psi/d^{m+2}\leq\hbar, 0\leq\psi<\Psi_{\rm{c}}\}\\&\ \ +\Pr\{\hbar\geq0, \cos\psi\leq\cos\Psi_{\rm{c}}\}\\
		&=\Pr\left\lbrace \cos\psi\leq\frac{\hbar d^{m+2}}{H_0}, 0\leq\psi<\Psi_{\rm{c}}\right\rbrace \\
		&+\Pr\{\cos\psi\leq\cos\Psi_{\rm{c}}, \hbar\geq 0\}\\
		&=F_{\cos\psi}\left(\frac{\hbar d^{m+2}}{H_0} \right) +F_{\cos\psi}\left(\cos\Psi_{\rm{c}}\right)\mathcal{U}(\hbar) .
		\end{aligned}
		\end{equation}
		where $\mathcal{U}(\hbar)$ is the unit step function; the support range of $F_H(\hbar)$ is $h_{\rm{min}}\leq\hbar\leq h_{\rm{max}}$. Then, the corresponding PDF can be obtained as follows:
		\begin{equation}
		\begin{aligned}
		\label{PDFH}
		f_H(\hbar)&=\frac{\partial}{\partial \hbar}F_H(\hbar)=\frac{d^{m+2}}{H_0}f_{\cos\psi}\left(\frac{\hbar d^{m+2}}{H_0} \right)+F_{\cos\psi}\left(\cos\Psi_{\rm{c}}\right)\delta(\hbar)\\
		&=\frac{c_{\rm{N}}}{h_{\rm{n}}}f_{\cos\psi}\left( \frac{\hbar}{h_{\rm{n}}}\right)+F_{\cos\psi}\left(\cos\Psi_{\rm{c}} \right)\delta(\hbar). 
		\end{aligned}
		\end{equation}
		where in the last equation, we define $h_{\rm{n}}\triangleq{H_0}/{d^{m+2}}$. The Dirac delta function comes out due to the discontinuity of CDF given in \eqref{CDFH} at $\hbar=0$. To ensure that the $\int_{h_{\rm{min}}}^{h_{\rm{max}}}f_H(\hbar){\rm{d}}\hbar=1$, the normalizing factor should be $c_{\rm{N}}=1$. 
		This completes the proof of \eqref{LOSPDF1}. 
	\end{proof}

	\bibliographystyle{./IEEEtran}
	\bibliography{./report}

\end{document}